\documentclass[journal,oneside]{IEEEtran}
\usepackage{mathtools}

\PassOptionsToPackage{bookmarks={false}}{hyperref}
\usepackage{graphicx}
\usepackage{lettrine}
\usepackage{booktabs}
\usepackage{amsmath}
\usepackage{enumerate}
\usepackage{bm}
\usepackage{graphicx}
\usepackage{epstopdf}
\usepackage{amsmath}
\usepackage{array}
\usepackage{amssymb}
\usepackage[numbers,sort&compress]{natbib}
\usepackage{placeins}
\usepackage{stfloats}
\usepackage{subfigure}
\usepackage{url}
\usepackage{epstopdf}

\usepackage[linesnumbered, ruled]{algorithm2e}

\usepackage{xcolor}
\usepackage{soul}

\newcolumntype{P}[1]{>{\raggedright\arraybackslash}p{#1}}

\ifCLASSINFOpdf
\else
\fi
  \usepackage[caption=false,font=footnotesize]{subfig}

\usepackage{tikz}
\usepackage{textcomp}

\newcommand\copyrighttext{%
  \footnotesize \textcopyright 2020 IEEE. Personal use of this material is permitted.
  Permission from IEEE must be obtained for all other uses, in any current or future 
  media, including reprinting/republishing this material for advertising or promotional 
  purposes, creating new collective works, for resale or redistribution to servers or 
  lists, or reuse of any copyrighted component of this work in other works. 
  DOI: 10.1109/TSG.2020.3032527}
\newcommand\copyrightnotice{%
\begin{tikzpicture}[remember picture,overlay]
\node[anchor=south,yshift=5pt] at (current page.south) {\fbox{\parbox{\dimexpr\textwidth-\fboxsep-\fboxrule\relax}{\copyrighttext}}};
\end{tikzpicture}%
}

\begin{document}
%
\title{Fault Detection for Covered Conductors \\ With High-Frequency Voltage Signals: \\ From Local Patterns to Global Features}

\author{
Kunjin Chen, Tom\'{a}\v{s} Vantuch, Yu Zhang, \emph{Member, IEEE}, Jun Hu, \emph{Member, IEEE}, and Jinliang He, \emph{Fellow, IEEE} 

\thanks{
This work was supported in part by the Natural Science Foundation of China under Grant 51720105004 and Grant 51921005, in part by the CZ.1.05/2.1.00/19.0389, Development of the ENET centre research infrastructure, and in part by a Hellman Fellowship and a Seed Fund Award from CITRIS and the Banatao Institute at the University of California.

K. J. Chen, J. Hu, and J. L. He are with the State Key Lab of Power Systems, Department of Electrical Engineering, Tsinghua University, Beijing 100084, P. R. of China. 

T. Vantuch is with the Centre ENET, V\v{S}B - Technical University of Ostrava, 17. listopadu 15, 708 33, Ostrava-Poruba, Czech Republic.

Y. Zhang is with the Department of Electrical and Computer Engineering, University of California, Santa Cruz, Santa Cruz, CA 95064, USA. 

(Corresponding author email: hejl@tsinghua.edu.cn).
} 
}

\maketitle

\maketitle
\copyrightnotice
\vspace{-10pt}



%

\maketitle
\begin{abstract}
The detection and characterization of partial discharge (PD) are crucial for the insulation diagnosis of overhead lines with covered conductors. With the release of a large dataset containing thousands of naturally obtained high-frequency voltage signals, data-driven analysis of fault-related PD patterns on an unprecedented scale becomes viable. The high diversity of PD patterns and background noise interferences motivates us to design an innovative pulse shape characterization method based on clustering techniques, which can dynamically identify a set of representative PD-related pulses. Capitalizing on those pulses as referential patterns, we construct insightful features and develop a novel machine learning model with a superior detection performance for early-stage covered conductor faults. The presented model outperforms the winning model in a Kaggle competition and provides the state-of-the-art solution to detect real-time disturbances in the field. 
\end{abstract}

\smallskip


%
\IEEEpeerreviewmaketitle

\section{Introduction}

Covered conductors have been widely used for medium voltage overhead lines in forested or dissected terrain areas because of their higher operational reliability and reduced land use \cite{kratky2018novel}. Compared with uninsulated overhead lines, the interphase touch of conductors or the contact with tree branches of covered conductors does not lead to an immediate short-circuit fault \cite{vantuch2016complex}. Nevertheless, a persistent contact with a tree branch may degrade conductor insulation over time and develop into a fault that hampers the normal operation of a power distribution system. Therefore, it is favorable to detect such faults to prevent full deterioration of the insulation system. A major indicator is the partial discharge (PD) activity induced by the fault \cite{mivsak2015testing, mivsak2017complex, kratky2018novel}.

Generally speaking, the PD pattern can be represented as an impulse component of the current or voltage signal generated by PD activity rooted from the insulation deterioration \cite{s2, s4}. The correct evaluation of PD patterns on the medium voltage overhead lines with covered conductors can be used for early detection of developing faults, thus improving the system's safety and reliability \cite{s5, s6, s7, mivsak2015testing}. Depending on the PD pattern evaluation from the current and voltage signals, two principles are currently used to detect covered conductor faults. The first principle is to evaluate the PD pattern as an impulse component of the current measured by Rogowski coil \cite{s5, s6, s7}. The second principle is to evaluate the PD pattern as an impulse component of the voltage signal measured by a capacitive divider \cite{mivsak2015testing}.

The data used in this work comes from a metering sensor that is based on a single layer coil wrapped around the covered conductor to acquire the voltage signal of electric stray field along the covered conductor \cite{mivsak2019towards}. High cost efficiency is an advantage of this solution compared to its competitors \cite{s6}, which leads to its broader deployment. However, to make it really work, an effective pattern recognition method is required to recognize faulty-related PD activities from its acquired signals. Several studies were brought to analyse the signal characters using fuzzy theory \cite{prilepok2017partial}, mathematical chaos \cite{lampart2018dynamical}, complex networks \cite{vantuch2016complex} or machine learning models \cite{mivsak2017complex}. To bring more attention on this problem, in 2018, a dataset which contained a large number of signal measurements was published on Kaggle, the world's largest data science collaboration platform \cite{kaggle}. The intent of the organizers was to attract worldwide researchers and data scientists to examine the signal data and to develop effective pattern recognition models for PD detection.

A successful fault detection model for covered conductors has to correctly handle the following factors:
\begin{itemize}
    \item The measurements are not synchronised with sinusoidal waveform shapes. Hence, accurate sine shape removal must be included.
    \item Deployment in nature-like environment introduces a significant and yet varying external background noise interference that interpolates false pulse patterns. A proper signal thresholding procedure is needed \cite{sriram2005signal}.
    \item A large number of various PD pattern types \cite{lampart2018dynamical} increases the problem's complexity, which requires a robust detection model.
    \item Naturally, a significantly higher number of fault-free signal samples results in a highly imbalanced dataset \cite{he2009learning}.
\end{itemize}

In this paper, we develop a novel fault detection model for covered conductors based on data-driven PD pattern exploitation. In the literature, signal processing and machine learning techniques have been widely used in the PD pattern detection in general \cite{sahoo2005trends,raymond2015partial} and were combined with various feature extraction techniques like wavelet decomposition \cite{ma2002interpretation}, recurrence quantification analysis \cite{kanakambaran2017identification}, fractal features \cite{satish1995can} or image processing. Two common types of PD pattern representations are phase-resolved PD patterns and individual PD pulse shapes. For phase-resolved PD patterns, the extracted features generally describe the distribution of quantities such as phase angles, charges, and counts \cite{lalitha2000wavelet, firuzi2018partial}. In addition, features can also be extracted by analyzing parameters related to pulse shapes \cite{alvarez2016clustering, mor2017new}. Extracted features are generally used as inputs for the machine learning models including random forest, support vector machines (SVM), gradient boosting machines \cite{kratky2018novel, mivsak2017complex}. 

The fault detection task is not about classifying the source of individual PD pulses, nor about classifying PD sources by using overall PD patterns accumulated over multiple cycles. The nature of early fault detection along with the complexity of signals collected by low-cost sensors from real environments \cite{kratky2018novel} requires that a good approach should be able to identify fault-related pulse patterns and aggregate information at the signal sample level. However, existing feature extraction methods mainly focus on individual pulses and aggregate the pulse-level features via simple statistics, which may be insufficient to reveal the fault-related information.

Our approach differentiates from existing methods mainly by its feature construction procedure. The major contribution of the present work is that we approach the pulse patterns in a data-driven manner and use clustering to bridge the gap between the locality of pulses and the desirability of building global features. The clustering results are not used for identifying the sources of pulses, but for constructing discriminative features. Different from features constructed using all pulses, cluster-specific features are not affected by pulses within other clusters. Novel features including template-matching degree and intra-cluster concentration degree are proposed. The results based on the same dataset and evaluation criteria from the Kaggle competition also make the model comparable and reproducible.

The organization of the paper is as follows. In Section II, detailed steps of the proposed approach are provided, which include signal pre-processing, pulse identification and feature construction. Section III covers the detailed numerical results and discussions. Finally, conclusions are drawn in Section IV.

\section{The Proposed Fault Detection Approach}

In this section, we first describe the dataset and the overall diagram of the proposed approach. Steps including signal pre-processing, pulse identification, and pulse clustering are implemented before the construction of features. The classifier used for fault detection and the evaluation metric are also introduced.

\subsection{Problem Description and Overview of the Approach}

\begin{figure}[!tb]
\centering
\includegraphics[width=7.5cm]{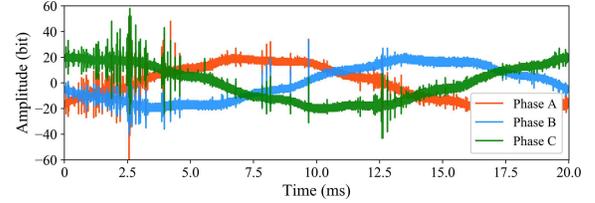}
\caption{An example of a three-phase signal labelled as faulty. The label is decided according to examination of a fault that was recorded after the timestamp of this signal sample, together with visual inspection of the pulses within the range from 0 to 5 ms.}
\label{signal_example}
\end{figure}

In Fig. \ref{signal_example}, an example of a three-phase high-sampling-rate voltage signal measured from a medium voltage overhead line is presented. The time span of the signals is 0.02 seconds and the sampling rate is 40 MHz (i.e., each signal has 800000 time steps). The fault detection task aims to determine whether developing covered conductor faults appear in the signals. To construct the training dataset, the one-cycle signals are labelled by a combination of expert observation and field examination \cite{mivsak2019towards}. For the faulty class, only the signals corresponding to the time range prior to the full development of faults are included. Therefore, a model that is able to classify faulty and non-faulty signals correctly can help prevent the covered conductors from being burned completely, which may lead to serious consequences like wildfires. A major challenge for the task is that some of the non-faulty signals appear to have faulty patterns \cite{mivsak2019towards}. It is worth noting that unlike lab experiments, no prior knowledge of fault-related PD patterns or background noise interference is available from the dataset. Therefore, a data-driven approach developed for this setting has a high potential to be generalizable to other measurement devices, interference sources, etc. More details of the fault detection task can be found in \cite{mivsak2015testing, kratky2018novel, mivsak2019towards}.

A total of 2904 three-phase signals are collected for the training dataset, while 2916 three-phase signals (the same as the \emph{private leaderboard} in the Kaggle competition) are used for testing. Out of the 8712 single-phase signals in the training dataset, 8187 signals (93.97\%) are non-faulty, which indicates that the dataset is quite imbalanced. Although faulty and non-faulty labels are manually assigned to each single-phase signal, we consider each three-phase signal as a whole and treat it as faulty if at least one of the single-phase signals is faulty. More details of the dataset can be found in \cite{mivsak2019towards}.

\begin{figure}[!tb]
\centering
\includegraphics[width=7.5cm]{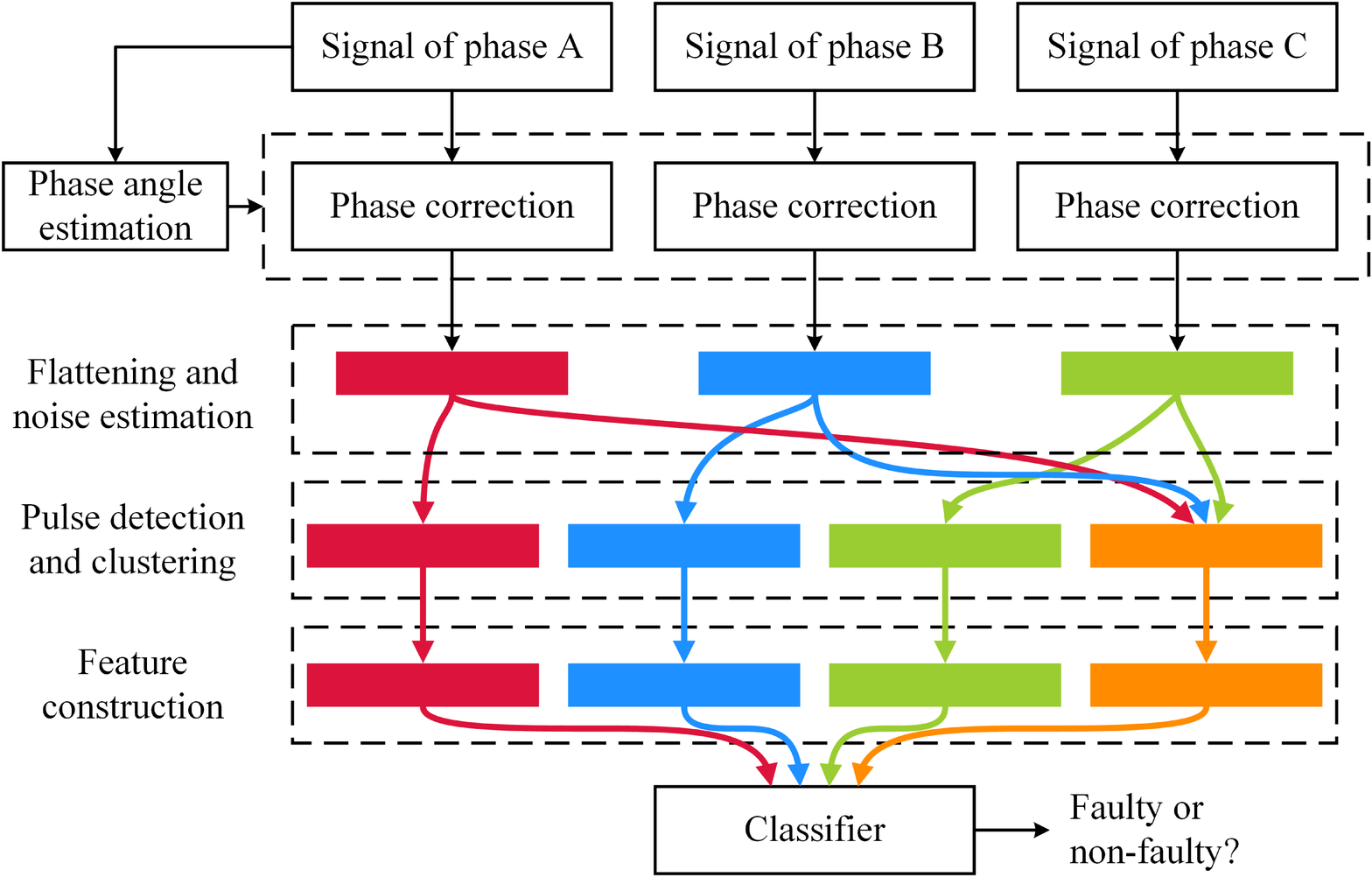}
\caption{The diagram of the proposed fault detection approach.}
\label{diagram}
\end{figure}

The diagram of the proposed approach is illustrated in Fig. \ref{diagram}, where the implementation steps are as follows:

\begin{itemize}

\item The three-phase signals first go through a phase correction process so that the signal of each phase is aligned to the same phase angle. 
\item The signals are flattened so that the pulses' amplitudes in the signals can be compared. A threshold is estimated for each signal to filter out the background noise. 
\item The pulses are detected by finding the points with the highest amplitudes within their neighbourhoods and the waveforms surrounding the pulses are clustered to reveal the waveform patterns. Clustering is performed for the three phases individually as well as combined.
\item Features are constructed based on the detected pulses and the clustering results. The features are then used to train a classifier that is able to distinguish between faulty and non-faulty signals.

\end{itemize}

\subsection{Signal Pre-processing}

The information related to PD is generally contained within the pulses in the signals. Thus, an important step before the implementation of machine learning models is to detect and analyze the pulses. Specifically, we propose a data-driven approach to reveal the characteristics of the pulses.

\begin{figure}[!tb]
\centering
\includegraphics[width=7.5cm]{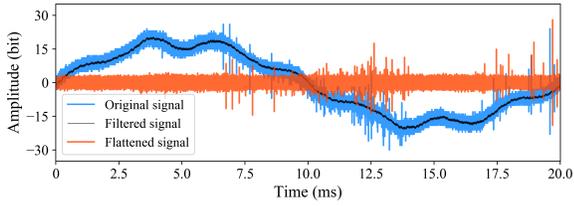}
\caption{An example of the signal flattening process.}
\label{flatten}
\end{figure}

Two steps of pre-prosessing are needed before we can detect pulses from the signals. The first step is phase angle correction and signal flattening, and the second step is noise level estimation. For phase angle correction, we use discrete Fourier transform (DFT) to extract the power frequency component in the signals and align the signals to have a phase angle of zero degree at the first time step. We then use the Savitzky-Golay filter \cite{savitzky1964smoothing} to fit the low frequency part of a signal, which is then subtracted from the signal so that the signal is flattened. Specifically, a value $y^*_j$ in the filtered signal can be calculated as 
\begin{equation}
y^*_j = \sum^{(m-1)/2}_{i=(1-m)/2}{C_iy_{j+i}},
\end{equation}
where $m$ is the window size and $C_i$ is the $i$th coefficient. The details for computing the coefficients can be found in \cite{schafer2011savitzky}. In Fig. \ref{flatten}, the process of filtering the signal with the Savitzky-Golay filter and signal flattening is illustrated.

\begin{figure}[!tb]
\centering
\includegraphics[width=7.5cm]{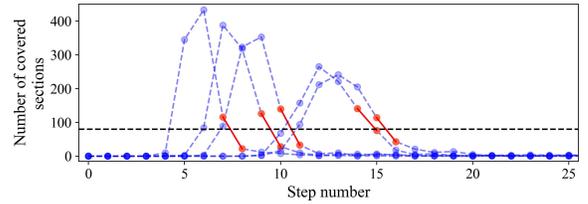}
\caption{An illustration of the noise estimation method with a threshold of 80. Five examples are randomly selected and the line sections crossing the threshold from top to bottom are highlighted.}
\label{noise_estimation}
\end{figure}

\begin{figure}[!tb]
\centering
\includegraphics[width=7cm]{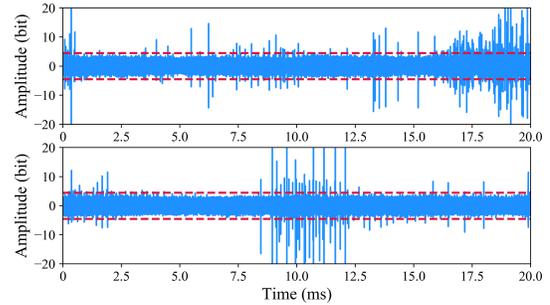}
\caption{Examples of estimated noise level on flattened signals.}
\label{noise_example}
\end{figure}

The method for noise level estimation is described in Algorithm \ref{algo:noise}. The inputs of the algorithm are the flattened signal $\bm{s}$, the number of signal sections $N_\mathrm{noise}$, the length of each signal section $L_\mathrm{noise}$, the threshold for the number of sections to cover $N_\mathrm{cover}$, the maximum noise level $C_\mathrm{max}$, and the incremental step size for noise level $C_\mathrm{step}$. For simplicity, we estimate one global noise level for each single-phase signal. An illustration of the noise estimation algorithm is illustrated in Fig. \ref{noise_estimation}. It can be seen that the threshold $N_\mathrm{cover}=80$ helps find a step number that has a low $m_i$ for each signal, such that the amplitudes covered in the section are not likely to belong to background noise. In Fig. \ref{noise_example}, we present examples of flattened signals together with estimated noise levels. It is observed that the estimated noise levels are able to cover the low-amplitude background noise without covering the high-amplitude pulses in the signals.

\begin{algorithm}[!tb]
\caption{Noise level estimation.}\label{algo:noise} 
\KwIn{$\bm{s}, N_\mathrm{noise}, L_\mathrm{noise}, N_\mathrm{cover}, C_\mathrm{max}, C_\mathrm{step}$} 
\KwOut{$a$} 
Sample $N_\mathrm{noise}$ sections ($\bar{\bm{s}}_1$ to $\bar{\bm{s}}_{N_\mathrm{noise}}$) with length $L_\mathrm{noise}$ equidistantly from $\bm{s}$ \\ 
\For{$i=1$ \KwTo $N_\mathrm{noise}$}
{
	$m_i \leftarrow \mathrm{max}(\mathrm{abs}(\bar{\bm{s}}_i))$
}

\For{$i=1$ \KwTo $C_\mathrm{max}/C_\mathrm{step}$} 
{ 
	$k \leftarrow 0$ \\
	\For{$j=1$ \KwTo $N_\mathrm{noise}$}
	{
		\If{$m_j < i \cdot C_\mathrm{step}$ \rm{and} $m_j >= (i-1) \cdot C_\mathrm{step}$}
		{
			$k \leftarrow k + 1$ \\
			\If{$k > N_\mathrm{cover}$}
			{
				$a \leftarrow i \cdot (C_\mathrm{step} + 1)$ \\
				\textbf{break}
			}
		}
	}
} 
return $a$ 
\end{algorithm}

\subsection{Pulse Identification}

Specifically, a point in a signal is considered a pulse if it has the highest absolute amplitude in its neighbourhood. A fast and robust way to locate the pulses in the signals is needed to facilitate the subsequent feature extraction and classification steps. Concretely, a two-stage pulse detection method is adopted, where the first stage finds a set of pulse candidates, and the second stage verifies if the candidates are valid pulses. The pulse detection method is described in detail in Algorithm \ref{algo:pulse}. Besides $\bm{s}$, the algorithm inputs also include the number of sections to equally split the signal $N_\mathrm{sort}$, the number of largest amplitude values to keep $N_\mathrm{top}$, the neighbourhood radius for masking $N_\mathrm{mask}$, the neighbourhood radius of local maximum $N_\mathrm{local}$, the estimated noise level $a$, and the threshold for magnitude comparison $C_\mathrm{mag}$.

\begin{figure}[!tb]
\centering
\includegraphics[width=7.5cm]{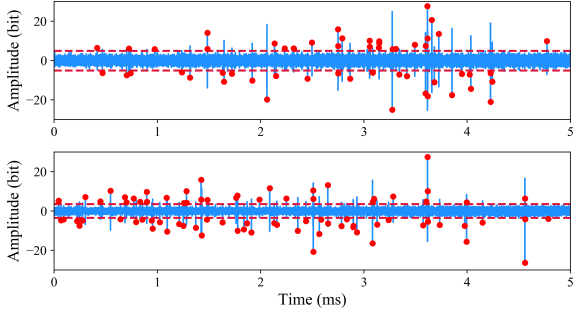}
\caption{Examples of detected pulses on flattened signals.}
\label{pulses}
\end{figure}

The first stage of the algorithm ensures that the pulses are not densely distributed within a small time section of the signals and that pulses of both high and low amplitudes can be selected. In the second stage, all pulse candidates collected from the first stage are verified to guarantees that the absolute amplitudes of the pulses are the maxima within their neighbourhoods. Examples of detected pulses can be seen in Fig. \ref{pulses}. In addition to the verification of amplitude, three conditions for relocation of pulses are described as $\mathrm{Cond}_i: \bm{s}[p-i] \bm{s}[p] < 0$ and $\bm{s}_\mathrm{abs}[p-i] > C_\mathrm{mag} \bm{s}_\mathrm{abs}[p], i=1,2,3$.
Relocation of the pulses is crucial for the construction of features based on the shapes of the waveforms surrounding the pulses. The conditions above are able to find high-amplitude points a few time steps before the pulse candidates.

\begin{algorithm}[!tb]
\caption{Pulse identification and processing.}\label{algo:pulse} 
\KwIn{$\bm{s}, N_\mathrm{sort}, N_\mathrm{top}, N_\mathrm{mask}, N_\mathrm{local}, a, C_\mathrm{mag}$} 
\KwOut{$P$} $\bm{s}_\mathrm{mask} \leftarrow \mathrm{abs}(\bm{s}), \bm{s}_\mathrm{abs} \leftarrow \mathrm{abs}(\bm{s}), P' \leftarrow \varnothing, P \leftarrow \varnothing$ \\
\For{$i=1$ \KwTo $3$}
{
	Divide $\bm{s}_\mathrm{mask}$ into $N_\mathrm{sort}$ sections ($\bar{\bm{s}}_1$ to $\bar{\bm{s}}_{N_\mathrm{sort}}$) \\
	\For{$j=1$ \KwTo $N_\mathrm{sort}$}
	{
		$P^\prime_j \leftarrow $ indexes of largest $N_\mathrm{top}$ values in $\bar{\bm{s}}_j$ \\
		Add $P^\prime_j$ to $P'$ \\
		\For{$p^\prime \in P^\prime_j$}
		{
			$\bm{s}_\mathrm{mask}[p^\prime - N_\mathrm{mask}: p^\prime + N_\mathrm{mask}] \leftarrow \bm{0}$\\
		}
	}
}

\For{$p \in P^\prime$}
{
	\If{$\bm{s}_\mathrm{abs}[p] = \mathrm{max}(\bm{s}_\mathrm{abs}[p-N_\mathrm{local} : p+N_\mathrm{local}])$}
	{
		\While{$\rm{Cond}_1$ \rm{is satisfied}}
		{
			$p \leftarrow p-1$ 
		}
		\uIf{$\rm{Cond}_2$ \rm{is satisfied}}
		{
			$p \leftarrow p-2$
		}
		\ElseIf{$\rm{Cond}_3$ \rm{is satisfied}}
		{
			$p \leftarrow p-3$
		}
		\If{$\bm{s}_\mathrm{abs}[p]>a$ \rm{and} $\bm{s}_\mathrm{abs}[p]<50$}
		{
			Add $p$ to $P$
		}
	}
}

return $P$ 

\end{algorithm}

\subsection{Pulse Clustering and Global Features Construction}
\label{section:sec2.4}

The previously-mentioned pulse detection method is able to locate the pulses with high absolute amplitudes within their neighbourhoods, but as we are more concerned about the waveforms surrounding the pulses, it is then of great importance to investigate the shapes of the waveforms. A straightforward way to compare the shapes of different waveforms is to align the waveforms with the pulses as the anchor point. That is, we normalize the waveforms surrounding the pulses by dividing the amplitude of the corresponding pulses and align the waveforms such that the pulses are located at the same position. Further, we expect that typical pulse waveform patterns can be revealed by clustering of the waveforms. The patterns will later serve to produce a set of finely tuned features describing the signal for the purpose of its classification. 

The clustering of pulses is carried out for individual phases as well as all phases combined. Up to $N_\mathrm{sample}$ pulses are randomly sampled from the signal of each phase for clustering so that the clustering process is not excessively influenced by signals with a large number of pulses. Specifically, k-means clustering with k-means++ initialization \cite{arthur2007K} is applied to the waveforms surrounding the identified pulses. Given a waveform $\bm{x}$, the length of the waveform is $N_\mathrm{before} + N_\mathrm{after} + 1$, where $N_\mathrm{before}$ and $N_\mathrm{after}$ are the number of time steps before and after the anchor point. After the clustering of pulse waveforms is performed, the centroids of the clusters can be calculated for further analysis.

The pulses detected in the signals and the clustering results cannot be directly used as inputs to machine learning models. Thus, it is necessary to build compact global features that can be used to distinguish between faulty and non-faulty signals. With the clustering results for individual phases and all phases combined, building features based on pulses in a specific phase (or all phases) or in a specific cluster becomes possible. Some traditional and trivial features include count of pulses, average height of pulses, and standard deviation (SD) of height of pulses. These features, however, are insufficient in reflecting the shapes of the pulses. Thus, two additional deep feature groups, namely, template-matching degree, and intra-cluster concentration degree, are introduced. Specifically, the features are constructed by calculating the rooted mean squared error (RMSE) between pulse waveforms and templates (for template-matching degrees) or cluster centroids (for intra-cluster concentration degrees).

Given the $i$th template $\bm{p}_i$, the corresponding template-matching feature $r_i$ is calculated as 
\begin{equation}
r_i = \frac{1}{N_\mathrm{pulse}}\sum\limits^{N_\mathrm{pulse}}\limits_{j=1}{\mathrm{RMSE}(\bm{x}_j, \bm{p}_i)},
\end{equation}
where $N_\mathrm{pulse}$ is the number of pulses detected in a three-phase signal. The templates are not necessarily cluster centroids but can be built based on the centroids. While the template-matching features are calculated using all pulses, the intra-cluster concentration degrees are obtained using pulses assigned to each cluster centroid. For each cluster centroid $\bm{\mu}_i$, the intra-cluster concentration feature $d_i$ is obtained by
\begin{equation}
d_i = \frac{\sum\limits_{\bm{x}_j:\mathrm{NC}(\bm{x}_j)=i}{\mathrm{RMSE}(\bm{x}_j, \bm{\mu}_i)}}{\sum\limits_{\bm{x}_j:\mathrm{NC}(\bm{x}_j)=i}{1}},
\end{equation}
where $\mathrm{NC}$ is the centroid assignment function given the cluster centroids. For phase-specific cluster centroids, the pulses are also limited in the phases. The features of all-phase cluster centroids use pulses from all phases.

\begin{figure}[!tb]
\centering
\includegraphics[width=6cm]{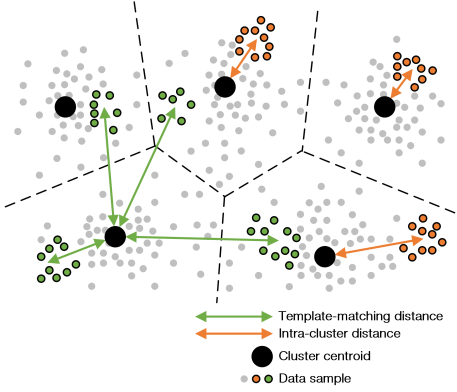}
\caption{An illustration of template-matching features (green dots) and intra-cluster concentration features (orange dots).}
\label{vonoroi}
\end{figure}

We use Fig. \ref{vonoroi} to demonstrate the concepts of template-matching and intra-cluster concentration features. Suppose the green data samples belong to a specific signal, then the template-matching features are calculated based on all the samples (indicated by the green arrows). Further, suppose the orange data samples belong to another signal, we can calculate the intra-cluster concentration features using the samples within each voronoi cell (indicated by the orange arrows). 

In the proposed approach, we construct two types of features, namely, \emph{all-pulse} features and \emph{cluster-specific} features. Specifically, all-pulse features are constructed using all pulses identified across three phases, while cluster-specific features are constructed with pulses within pulse clusters (both for individual phases and all phases combined). As each pulse is assigned to one of the clusters, features constructed for a specific cluster are not affected by pulses in other clusters. In addition, template-matching features are built for all-pulse features only and intra-cluster concentration features are exclusive for cluster-specific features. Pulse count, average height, and SD of height features are used for both types of features. When all the features are constructed, we put them together and build classifiers based on the features.

\subsection{Classifier and Evaluation Metric}

In this paper, we use light gradient boosting machine (lightGBM) \cite{ke2017lightgbm} as the classifier. The lightGBM model is an efficient implementation of gradient boosting decision tree (GBDT) \cite{friedman2001greedy}, a gradient boosting framework built upon tree-based learning algorithms. Specifically, the GBDT model is an addictive ensemble of decision trees that are trained in a sequential manner, and each tree is trained based on the negative gradients of the loss function. The implementation of lightGBM is based on XGBoost \cite{chen2016xgboost}, which optimizes the loss function using second-order approximation. The lightGBM model is suitable for the features we have constructed because of the following reasons:
\begin{enumerate}
\item Tree-based models are in general insensitive to the values of the features, thus only the raw features are needed. Specifically, the values of a feature are ordered and put into bins by building a histogram. For each split, values within bins on two sides of the split point are separated into two new leaves.
\item The lightGBM model automatically decides which feature to use when splitting a leaf node. Thus, the importance of a feature can be reflected by the number it is used or the gain of the feature. 
\end{enumerate}
Specifically, binary-crossentropy is used as the loss function for lightGBM for the binary classification task in this work. As non-faulty samples account for over 93\% of the training dataset, the training of the classification model may be hampered by this imbalance. One effective method to tackle class imbalance is over-sampling the minority class (faulty) near the borderline between two classes, which can lead to an improved decision boundary with a mild increase in dataset size. In this work, borderline synthetic minority oversampling technique (SMOTE) with SVM is used to create synthetic faulty samples. Specifically, SMOTE-SVM \cite{nguyen2011borderline} first trains an SVM on the original dataset to approximate the borderline area with support vectors. New faulty samples are synthesized near support vectors for the faulty class so that the desired ratio of the number of faulty samples to the number of non-faulty samples, $\alpha$, is achieved (interested readers are referred to \cite{nguyen2011borderline} for the details). The SMOTE-SVM is used as a final ingredient for our proposed approach.

The Matthews correlation coefficient (MCC) is used to evaluate the fault classification performance of different approaches, and it is defined as
\begin{equation}
\begin{split}
& MCC = \\
& \frac{TP\times TN-FP\times FN}{\sqrt{(TP+FP)(TP+FN)(TN+FP)(TN+FN)}},
\end{split}
\end{equation}
where $TP$ is the true positive rate (the rate samples predicted as positive are indeed true samples, similarly hereinafter); $TN$ the true negative rate; $FP$ the false positive rate; and $FN$ the false negative rate. The MCC is considered a good evaluation metric for binary classification tasks on imbalanced datasets \cite{chicco2020advantages}.

\section{Results and Discussion}

In this section, we describe the implementation details of the proposed approach and present the results for fault detection. The performance of the proposed approach is enhanced by analyzing feature importance values. Additionally, a discussion on the number of clusters is provided.

\subsection{Implementation Details for Signal Pre-processing and Pulse Identification}

The implementation details of the methods for signal pre-processing and pulse identification are elaborated as follows:
\begin{enumerate}
\item Signal flattening: the Savitzky-Golay filter uses a window size of 99 and a polynomial order of 3.
\item Noise level estimation: $N_\mathrm{noise}$ and $L_\mathrm{noise}$ are set to 1000 for the sampling of signal sections. The threshold $N_\mathrm{cover}$ is set to 80. Considering the possible range of noise level, $C_\mathrm{max}$ is set to 15 and the step size $C_\mathrm{step}$ is $-$0.5.
\item Pulse identification and processing: We set $N_\mathrm{sort}$ and $N_\mathrm{top}$ to 20 and 100, respectively. The values for $N_\mathrm{mask}$ and $N_\mathrm{local}$ are 50 and 25. The magnitude coefficient $C_\mathrm{mag}$ is set to 0.5.
\end{enumerate}

\subsection{Analysis of Pulse Patterns and Clustering of pulses}
\label{section:sec3.2}

\begin{figure}[!tb]
\centering
\subfigure[]{
\centering
\includegraphics[width=7cm]{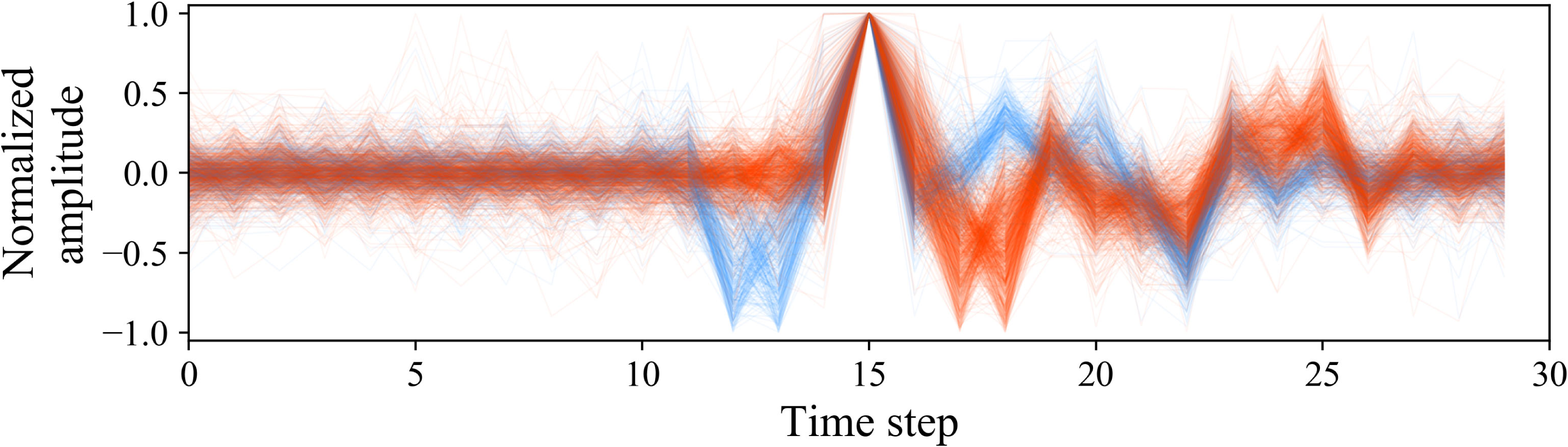}
}%
\quad
\subfigure[]{
\includegraphics[width=7cm]{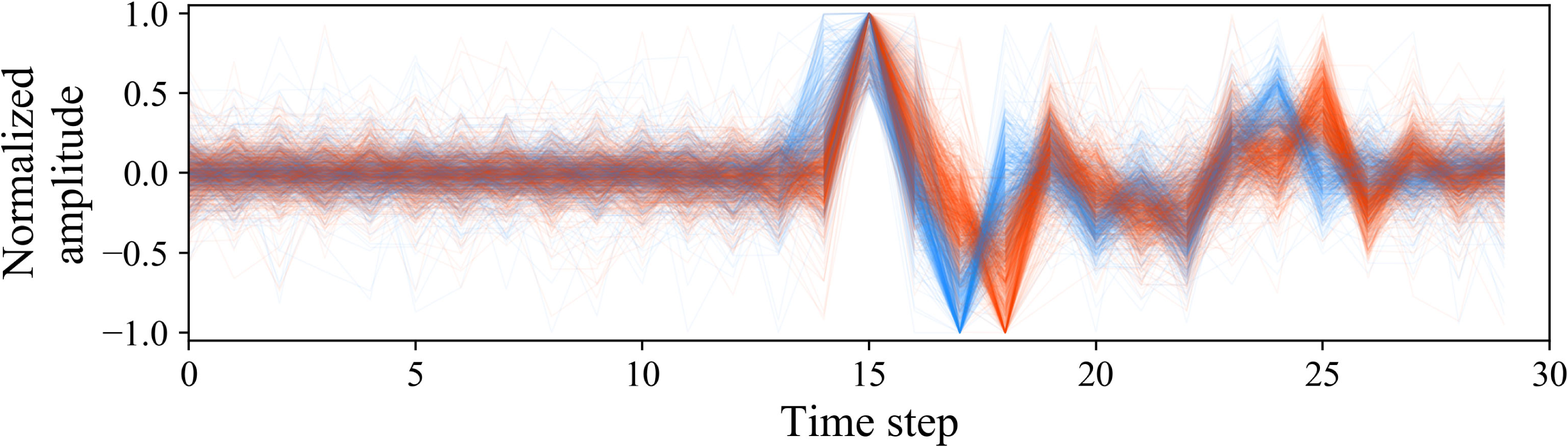}
}%
\caption{Two clusters of the waveforms of pulses for phase B before and after modifying the anchor points: (a) before and (b) after.}
\label{waves_after}
\vspace{-3pt}
\end{figure}

\begin{figure}[!tb]
\centering
\includegraphics[width=7.5cm]{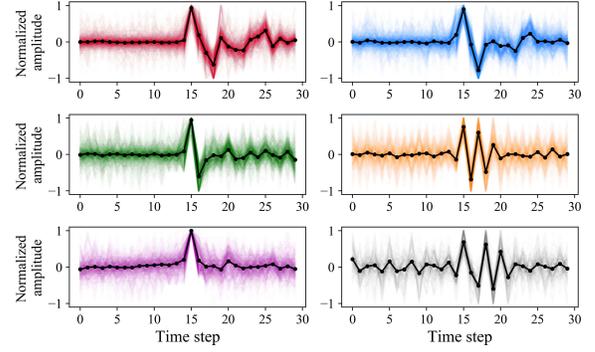}
\caption{Centroids for the waveforms of different clusters for phase B when $k=6$. Random waveform samples of the clusters are also illustrated.}
\label{centroids_B}
\vspace{-3pt}
\end{figure}

\begin{figure}[!t]
\centering
\includegraphics[width=6cm]{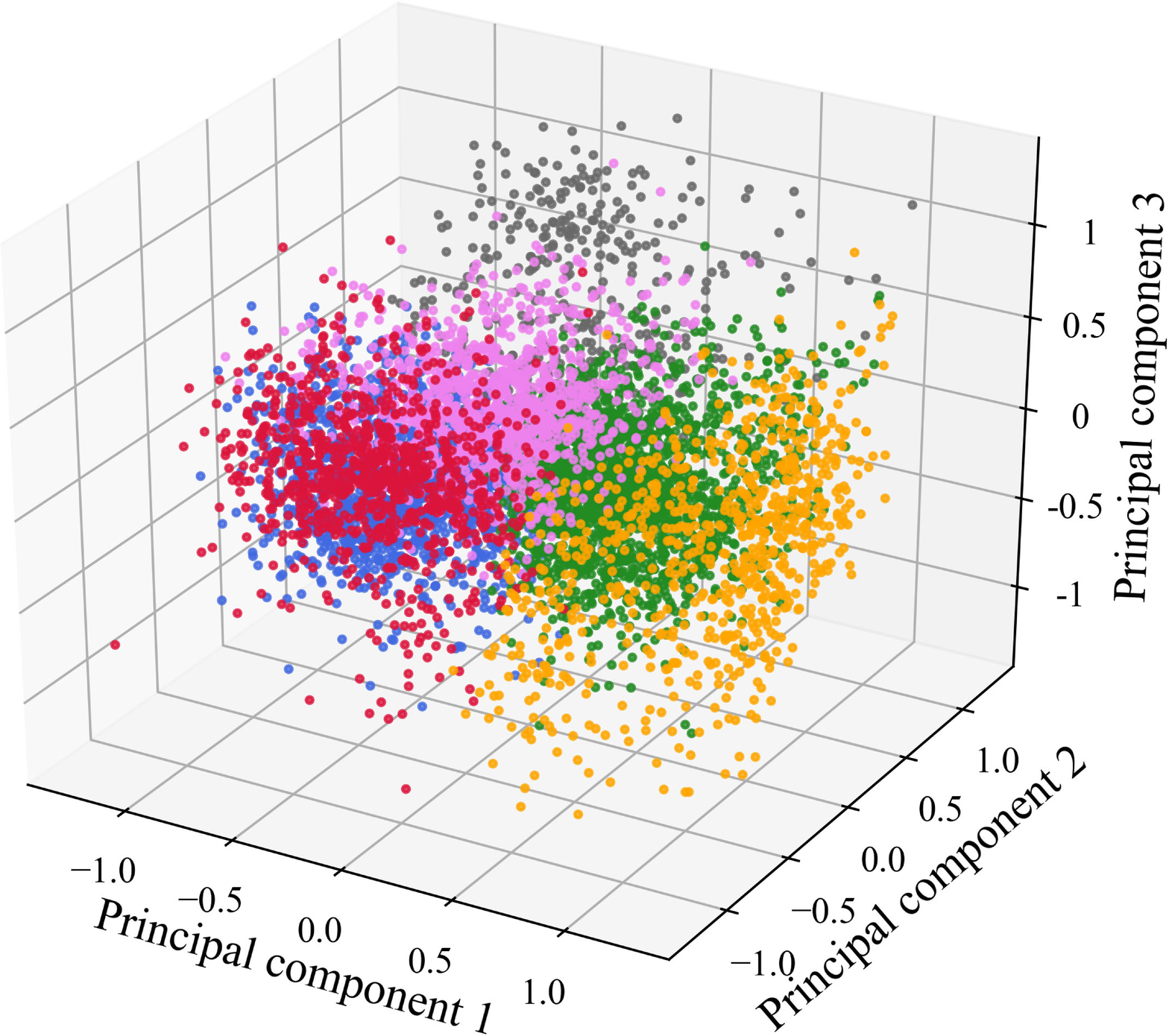}
\caption{Visualization of the clustering result for phase B using three principal components.}
\label{scatter}
\vspace{-3pt}
\end{figure}

\begin{figure}[!tb]
\centering
\includegraphics[width=7.5cm]{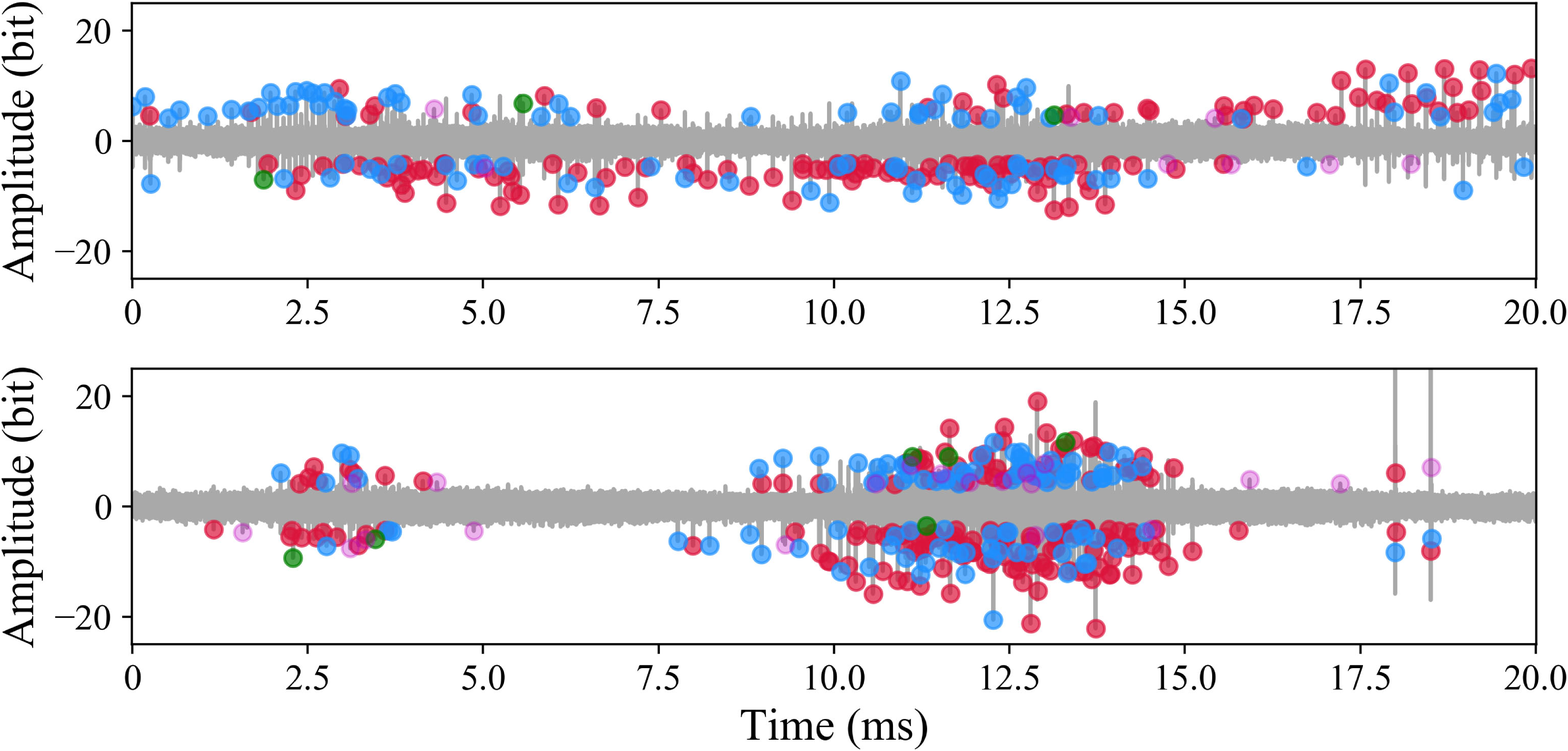}
\caption{Examples of flattened signals for phase B that are faulty. The detected pulses are marked with colors corresponding to the types in Fig. \ref{centroids_B}.}
\label{map_faulty}
\vspace{-3pt}
\end{figure}

\begin{figure}[!tb]
\centering
\includegraphics[width=7.5cm]{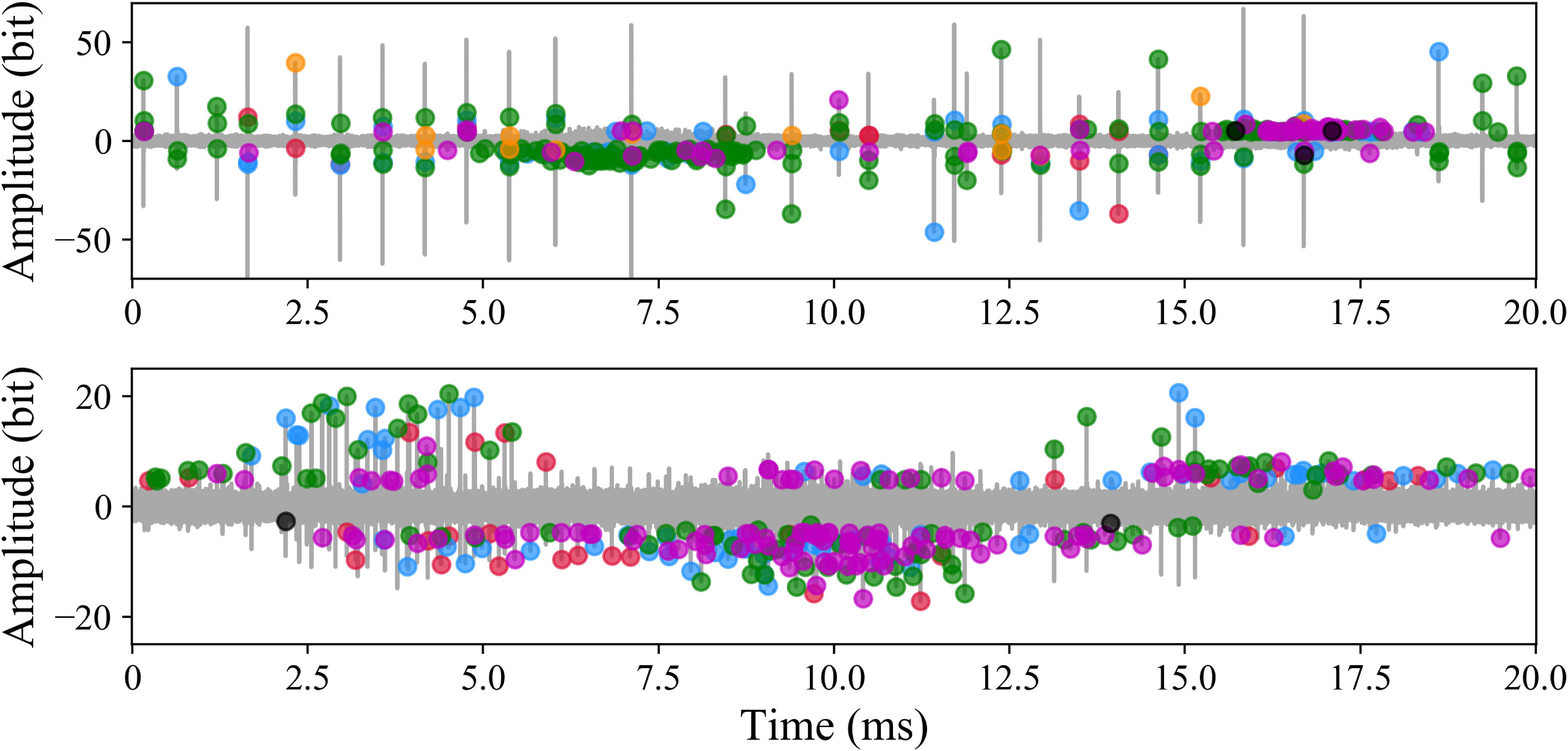}
\caption{Examples of flattened signals for phase B that are not faulty. The detected pulses are marked with colors corresponding to the types in Fig. \ref{centroids_B}.}
\label{map_non_faulty}
\vspace{-3pt}
\end{figure}

\begin{figure}[!tb]
\centering
\subfigure[]{
\centering
\includegraphics[width=7cm]{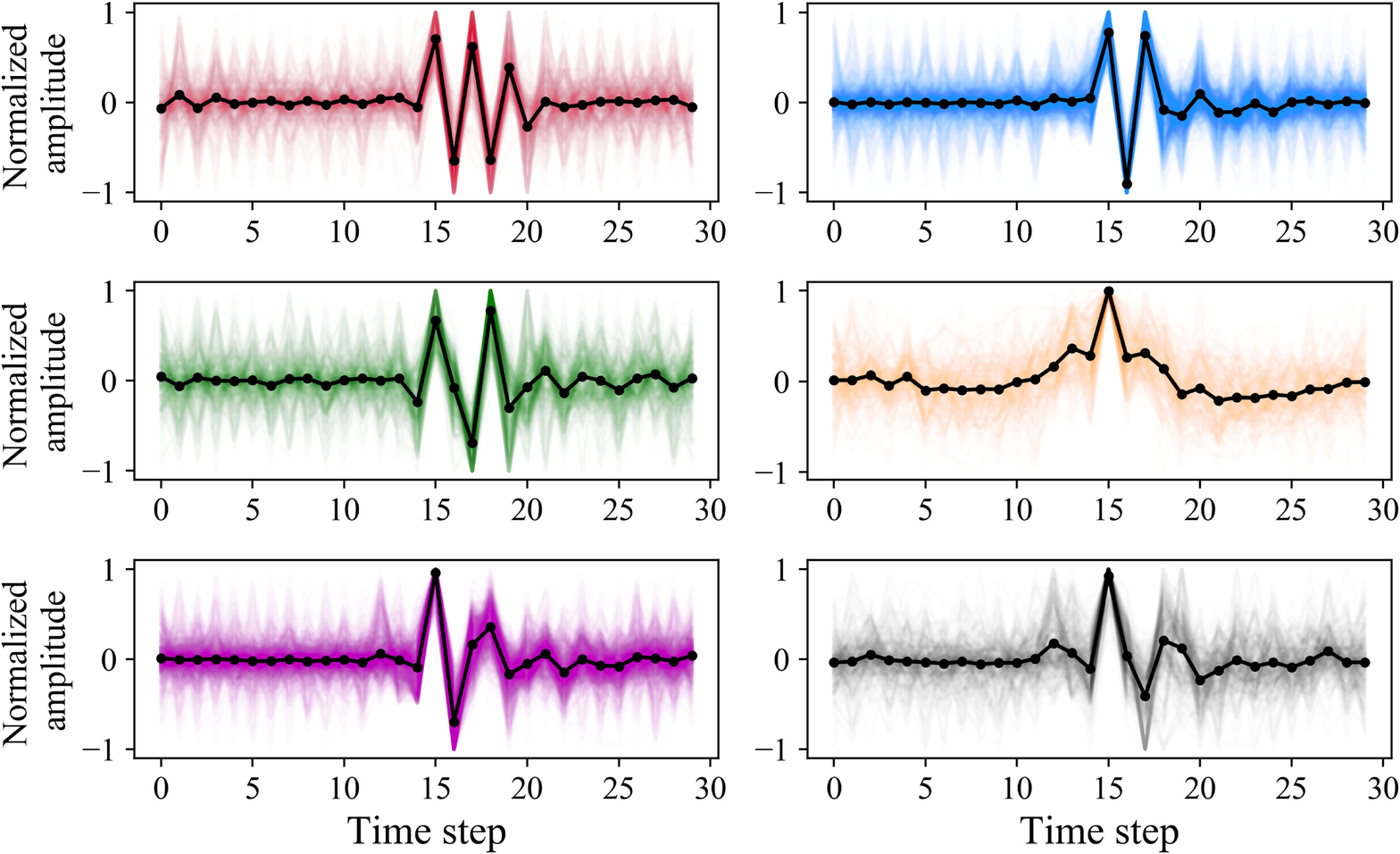}
}%
\quad
\subfigure[]{
\centering
\includegraphics[width=7cm]{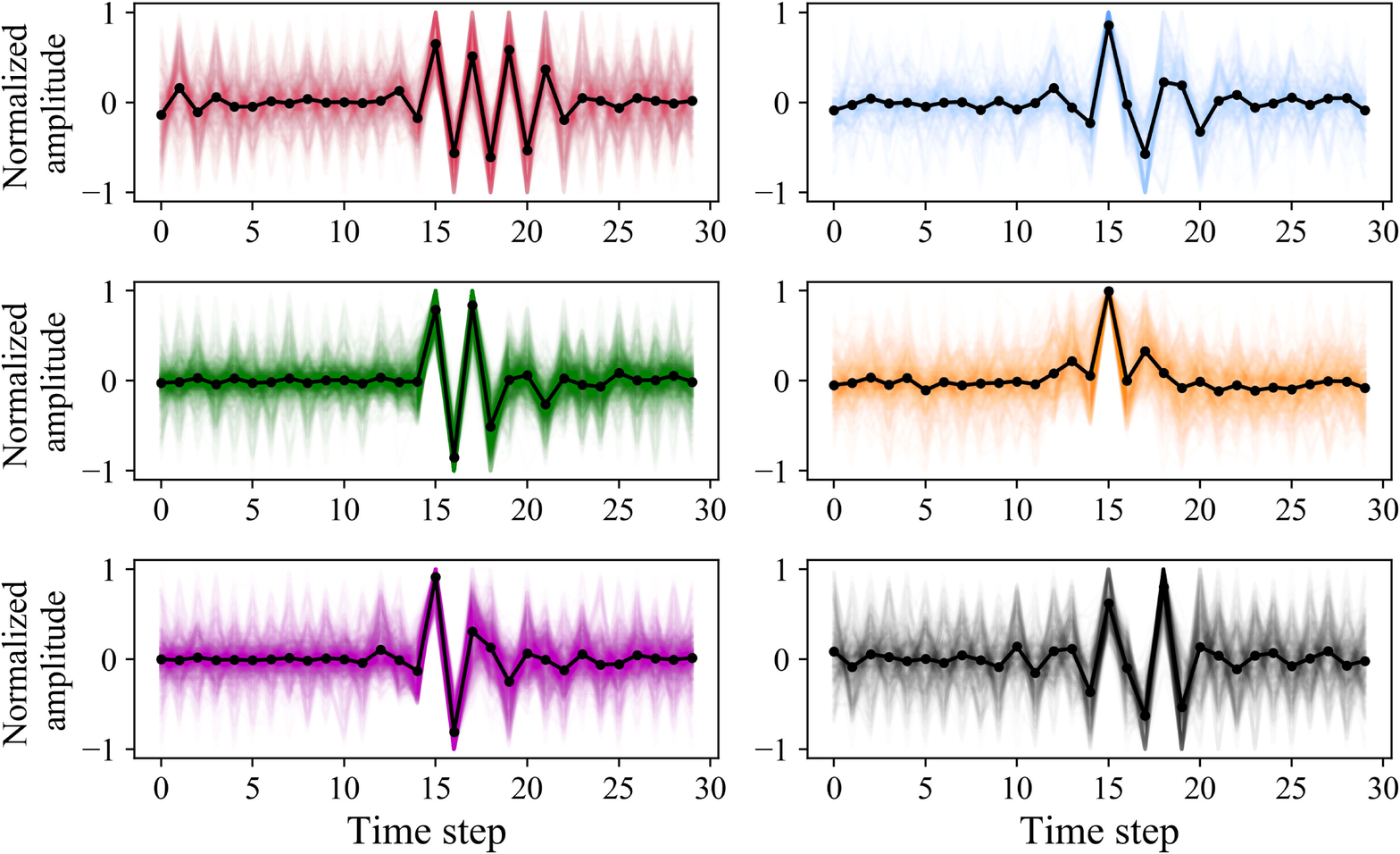}
}%
\caption{Centroids for the waveforms of different clusters for (a) phase A and (b) phase C when $k=6$.}
\label{centroids_AC}
\vspace{-3pt}
\end{figure}

\begin{figure}[!tb]
\centering
\subfigure[]{
\centering
\includegraphics[width=7cm]{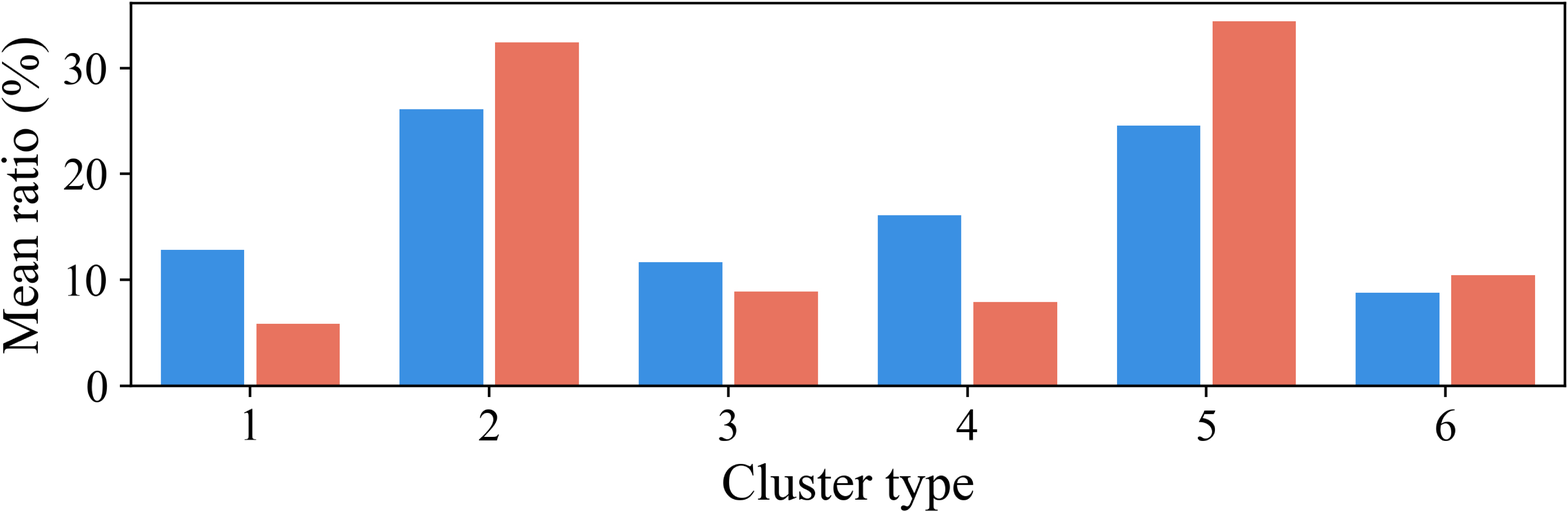}
}%
\quad
\subfigure[]{
\centering
\includegraphics[width=7cm]{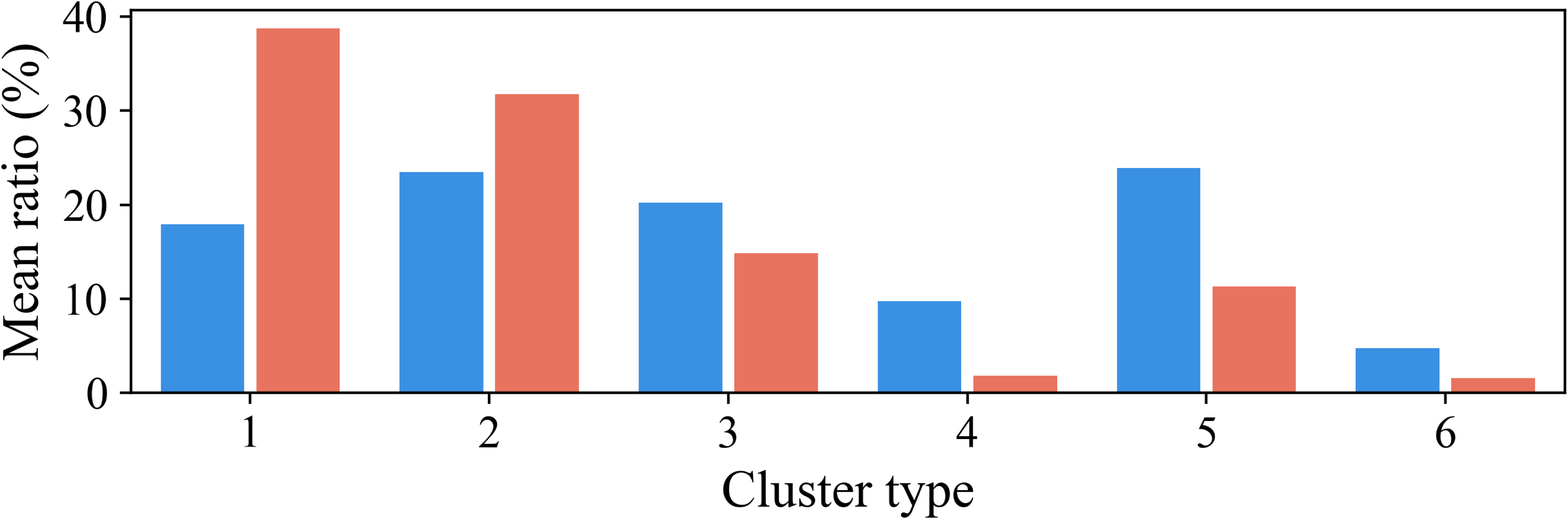}
}%
\quad
\subfigure[]{
\centering
\includegraphics[width=7cm]{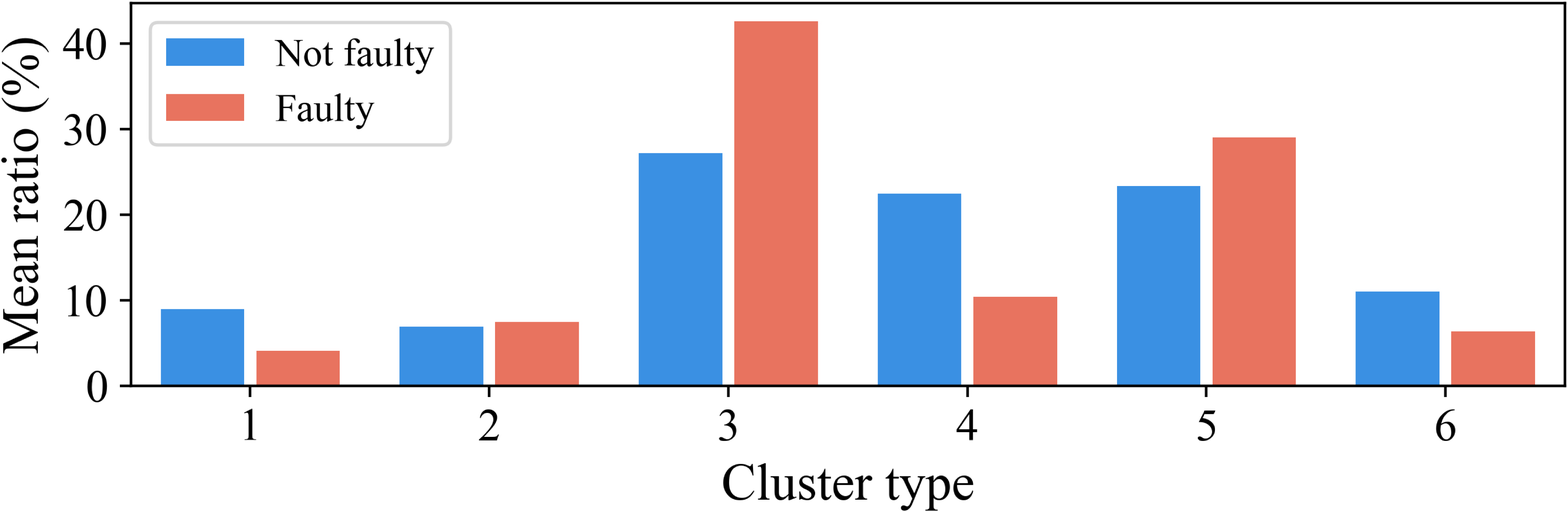}
}%
\caption{The average ratio of pulse counts in each cluster to the number of pulses in faulty and non-faulty signals for (a) phase A, (b) phase B, and (c) phase C.}
\label{ratios}
\vspace{-3pt}
\end{figure}

\begin{figure*}[!tbh]
\centering
\includegraphics[width=15cm]{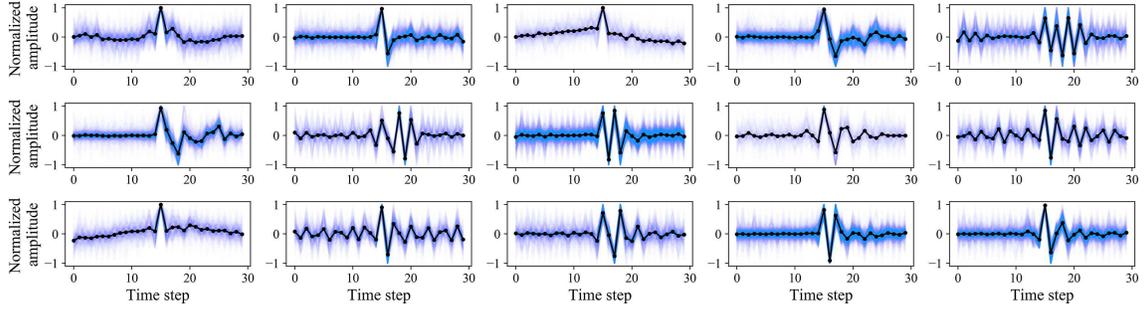}
\caption{Centroids for the waveforms of the detected pulses of different clusters for all phases when $k=15$. The ordering of the types is from left to right, then from top to bottom.}
\label{centroids_all}
\vspace{-3pt}
\end{figure*}

\begin{figure}[!t]
\centering
\subfigure[]{
\centering
\includegraphics[width=4cm]{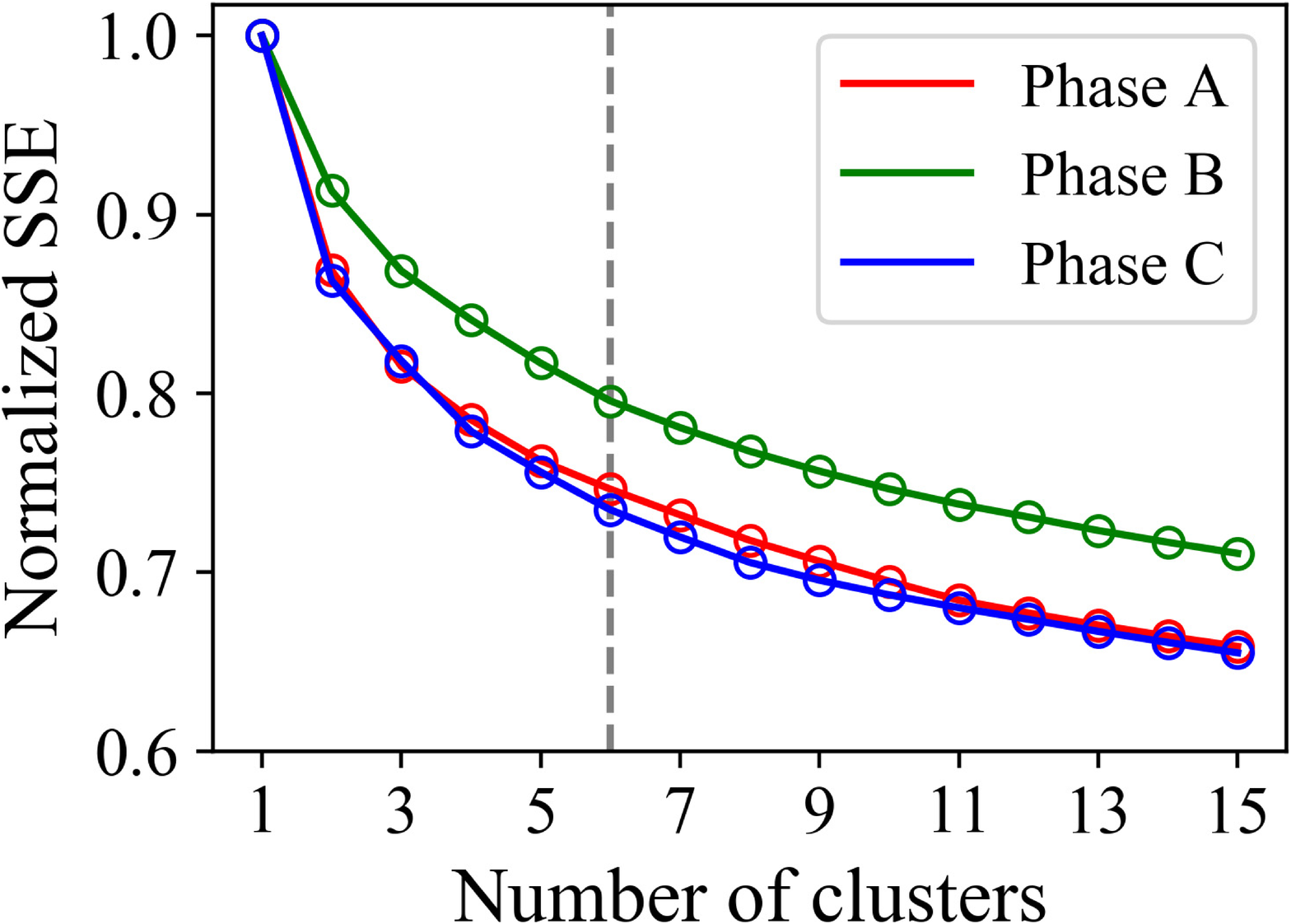}
}%
\quad
\subfigure[]{
\includegraphics[width=4cm]{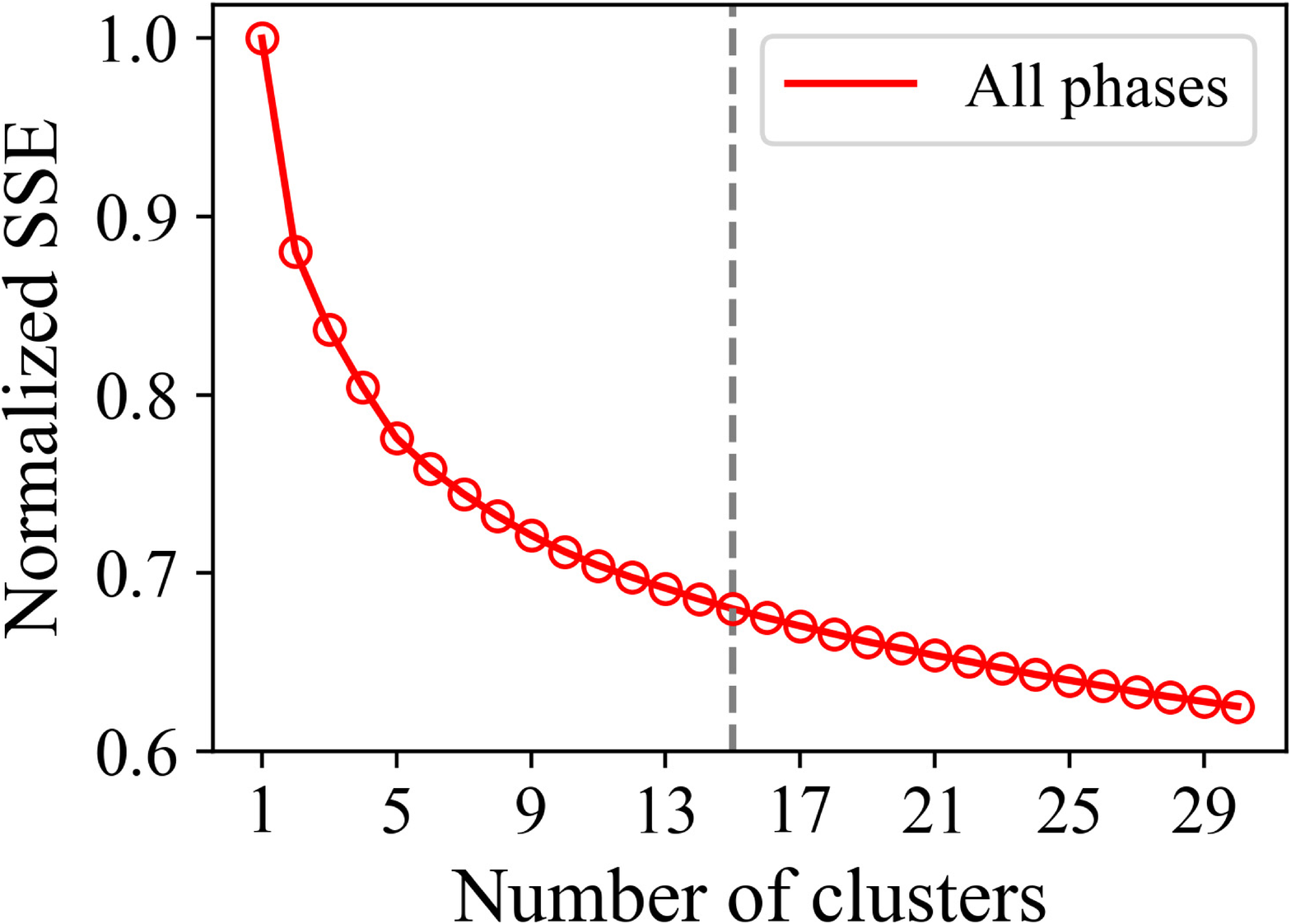}
}%
\caption{The values of normalized SSE for (a) individual phases and (b) all phases.}
\label{SSE}
\vspace{-3pt}
\end{figure}

\begin{figure}[!t]
\centering
\includegraphics[width=8cm]{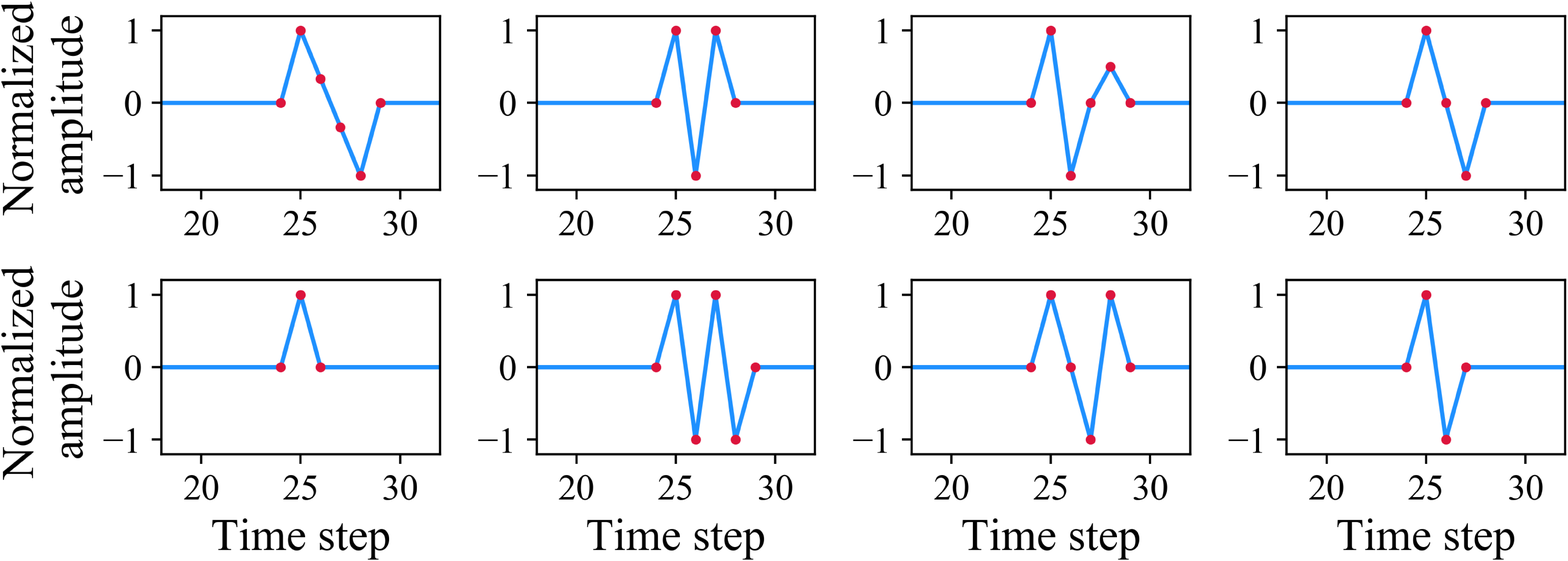}
\caption{The templates used for constructing the template-matching features.}
\label{templates}
\vspace{-3pt}
\end{figure}

Clustering of waveforms surrounding the pulses plays an important role in the proposed approach. In this subsection, we first use clustering of pulse waveforms from phase B to explain why the conditions in Algorithm \ref{algo:pulse} are added. Specifically, the inspection of the clustering results indicate that relocation of the pulses is needed to reveal the characteristics of the waveforms surrounding pulses. After the pulses are identified, we perform clustering to the pulse waveforms from individual phases as well as from all three phases, producing four groups of clustering results that are further exploited to construct features. Specifically, the clustering results are obtained with $N_\mathrm{sample}=100$. 

We use the waveforms taken from phase B signals that are faulty to illustrate the pulse waveform inspection procedure. As a first step, the 30-time-step waveforms ($N_\mathrm{before}=15$ and $N_\mathrm{after}=14$) are clustered into three groups using the k-means algorithm with $k=3$. Waveforms of two clusters are shown in Fig. \ref{waves_after} (a) with different colors, and it is observed that both clusters have two sub-modes within them. We shift the anchor points of the waveforms in phase B two or three steps ahead if either of the two time steps has an amplitude lower than $-$0.5. It is clearly observed in Fig. \ref{waves_after} (b) that the modification of the anchor points helps reveal the two types of waveforms that have an opposite pulse two or three steps after the anchor point. In addition, a similar modification of anchor points for waveforms that has an amplitude lower than $-$0.5 one time step before the anchor point is implemented. As multiple pulses with opposite signs may exist, the modification process can repeat for more than once.

The previous analysis is only for waveforms from faulty signals, which does not cover the complete characteristics of the different waveform types in phase B. A larger value of k can be selected to discover more types of waveforms. Here, a simple heuristics is used to determine the value of $k$. As The two waveform types in Fig. \ref{waves_after} (b) are quite similar to each other, waveforms of the two types are prone to be found within the same cluster if waveforms from non-faulty signals are included and the value of $k$ is small. Thus, we increase the value of $k$ and set it to the first value that is able to separate waveforms from the two types. Specifically, $k=6$ satisfies the criterion and we use the same value for the clustering of waveforms for phase A and C.

The six waveform types for phase B signals are illustrated in Fig. \ref{centroids_B}. The centroids of the clusters are plotted with black lines. Apart from the two types discussed earlier, the four remaining types are also quite distinctive. A visualization of the clustering result for phase B is shown in Fig. \ref{scatter}. Principal component analysis (PCA) \cite{wold1987principal} is used to reduce the dimensionality of the waveforms and the first three principal components are chosen. For ease of presentation, 15\% of the pulse waveforms are randomly sampled for the visualization. In addition, we use a few examples to show how the different waveform types can be used to distinguish between faulty and non-faulty signals. In Fig. \ref{map_faulty} and Fig. \ref{map_non_faulty}, we use dots with different colors to indicate the location of different types of waveforms in the signals. The colors for the dots are identical for the waveforms of different types in Fig. \ref{centroids_B}. It can be observed from the figures that faulty and non-faulty signals differ in terms of the types of waveforms surrounding pulses and the density of pulses. 

The clustering of pulse waveforms for phase A and C can be implemented in a similar way, and the centroids of the clusters for the two phases are shown in Fig. \ref{centroids_AC}. Although the orders of the centroids for the two phases are not matched, it can be observed that there is a one-to-one relationship between the centroids. It is worth pointing out that the pulse waveforms in each cluster appear in both faulty and non-faulty signals. To better illustrate this, the average ratios of pulse counts in the clusters to the number of pulses in faulty and non-faulty signals for each phase are given in Fig. \ref{ratios}. It can be seen that although some clusters account for higher proportions in faulty signals, these clusters still share relatively large proportions in non-faulty signals. Therefore, it is more important to use the clustering results for the construction of distinctive features instead of using them as fault indicators directly. In addition, it is impractical to classify the source of all the pulses appearing in the field signals. The data-driven approach based on clustering can provide clues about early-stage faulty conditions without explicitly identifying the fault-related pulses.

In addition to phase-specific clustering of waveforms, the clustering of waveforms from all phases is also conducted. In Fig. \ref{centroids_all}, the waveforms of the clusters with $k=15$ are presented. For most of the centroids in the figure, it is not difficult to find similar centroids for individual phases.

The sum of squared error (SSE) (i.e., the sum of squared distance from each pulse to its centroid) can be used to choose proper values of $k$ for k-means clustering. In Fig. \ref{SSE}, we plot the values of SSE for the several cases with different values of $k$. We normalize the SSE values so that the first value is always 1. A simple method for choosing $k$ is picking a $k$ so that the slope of the curve after the $k$th point becomes relatively small and stable. While it is hard to determine optimal values of $k$ from the curves, we can see that $k=6$ for individual phases and $k=15$ for all phases (dashed lines) roughly satisfy the criterion of the heuristic method.

\subsection{Description of Features and Classifier}
\label{section:sec3.3}

As is introduced in \ref{section:sec2.4}, we construct all-pulse and cluster-specific features separately. The descriptions for the all-pulse features are provided in Table \ref{feat_global}. The features in Table I are referenced from the Kaggle's 1st place solution, and the templates for template-matching features are shown in Fig. \ref{templates}. The length of the templates is 50 time steps. Initially, only the first template in Fig. \ref{templates} is introduced in the Kaggle's 1st place solution. We add 7 more features to this feature group based on the centroids illustrated in Fig. \ref{centroids_all}. In Table \ref{feat_cluster}, we describe the cluster-specific features in detail. Each feature group has 33 features, 15 of which are for all phases combined and 18 of which are for individual phases (6 for each phase). 

Similar to the implementation in the Kaggle's 1st place solution, a total of 125 lightGBM models are trained with 25 different random seeds and 5-fold cross-validation \cite{yadav2016analysis}. Generally speaking, 5-fold cross-validation means that we split the training dataset into 5 disjoint parts and leave one of the parts out at a time as the validation set. In order to compare with the Kaggle's 1st place solution, we follow its cross-validation scheme that uses 60\% of the data for training, 20\% for validation, and the remaining 20\% for testing. After determining the features and the hyper-parameters, five models are trained in the 5-fold manner. With a different random seed, the split of the training dataset changes so that more models can be trained to facilitate ensemble learning \cite{dietterich2000ensemble}. The final logits are calculated by averaging the logits of all models and the threshold for the logits is estimated by training samples.

\begin{table*}[tbh]
\renewcommand\arraystretch{1.2}
\centering  
\captionsetup{justification=centering}
\caption{Description of All-Pulse Features} 
\begin{tabular}{l l c}
\toprule[1.5pt]
Feature name & Description of the feature & Number of features \\
\midrule[0.75pt]
Pulse count  & Number of pulses for quadrants one and three, quadrants two and four, and all quadrants &  3\\
Average height  &  Average pulse height for quadrants one and three         &  1\\
SD of height &  SD of pulse heights for quadrants one and three  &  1         \\
Template-matching degree  & Average RMSE between the waveforms and templates for quadrants one and three &  8  \\
\bottomrule[1.5pt]
\end{tabular}
\label{feat_global}
\end{table*}

\begin{table*}[tbh]
\renewcommand\arraystretch{1.2}
\centering  
\captionsetup{justification=centering}
\caption{Description of Cluster-Specific Features Based on pulses of Different Clusters} 
\begin{tabular}{l l c}
\toprule[1.5pt]
Feature name & Description of the feature & Number of features\\
\midrule[0.75pt]
Pulse count  &  Number of pulses for each cluster &  33 \\
Average height  &  Average pulse height for each cluster   &  33 \\
SD of height &  SD of pulse heights for each cluster  &  33         \\
Intra-cluster concentration degree  &  Average RMSE between the cluster centroids and waveforms in the clusters  &  33   \\
\bottomrule[1.5pt]
\end{tabular}
\label{feat_cluster}
\end{table*}

The experiments of the proposed approach are conducted in Python 3.6 using CPUs. With a single thread, the average time lengths for feature extraction (three phases) and inference of 125 LightGBM models are 0.0156 and 0.0055 seconds, respectively. Using multiple threads for calculation further reduces the computation time. For instance, it takes less than 0.0008 seconds for the inference of the LightGBM models when 8 threads are used. Hence, our efficient approach is suitable for real-time fault detection applications.

\subsection{Fault Detection Results and Comparison}
\label{section:sec3.4}

\begin{table}[!t]
\renewcommand\arraystretch{1}
\centering  
\captionsetup{justification=centering}
\caption{Comparison of the MCC Values for Different Approaches} 
\begin{tabular}{p{2.5cm} p{1.2cm} p{1.2cm} p{1.2cm}}
\toprule[1.5pt]

Model & MCC & Precision & Recall\\
 
\midrule[0.75pt]
Reference \cite{mivsak2017complex} & 0.589 & 0.573 & 0.643\\
Kaggle 5th place     & 0.697  & - & -\\ 
Kaggle 2nd place     & 0.715  & - & -\\ 
Kaggle 1st place     & 0.719  & 0.616 & \textbf{0.867} \\ 
Proposed             & \textbf{0.735} & \textbf{0.669} & 0.841\\

\bottomrule[1.5pt]
\end{tabular}
\label{compare}
\end{table}

We compare the performance of the proposed model with other approaches. Spcecifically, the 1st place solution\footnote{\url{https://www.kaggle.com/mark4h/vsb-1st-place-solution}} is also based on features constructed on pulses, but the majority of the features are pulse counts and statistics of pulse heights. The 2nd place solution\footnote{\url{https://www.kaggle.com/c/vsb-power-line-fault-detection/discussion/86616#latest-501584}} is a combination of CatBoost models \cite{prokhorenkova2018catboost} and recurrent neural network models. The 5th place solution\footnote{\url{https://www.kaggle.com/c/vsb-power-line-fault-detection/discussion/85170#latest-500367}} combines lightGBM models with recurrent neural networks and convolutional neural networks. The approach in \cite{mivsak2017complex}, which uses a series of traditional features including count, width, and height of pulses, is also implemented as a benchmark. In table \ref{compare}, the metrics of the different models are presented. It is seen in the table that the proposed approach outperforms the benchmark as well as the top solutions for the competition on MCC. The metrics of precision $\frac{TP}{TP+FP}$ and recall $\frac{TP}{TP+FN}$ are also shown in Table \ref{compare}. It is observed in the table that the proposed approach has much higher precision. Specifically, there are 377 faulty signals in the test dataset, 327 out of the 531 signals predicted as faulty are actually faulty for the Kaggle's 1st place solution, and 317 out of the 474 signals predicted as faulty are true positives for the proposed approach. In a nutshell, the proposed approach is less likely to produce false positive predictions while the recall rates of the two approaches are roughly at the same level.

\subsection{Performance Enhancement With Feature Analysis}

\begin{figure*}[!tb]
\centering
\includegraphics[width=18cm]{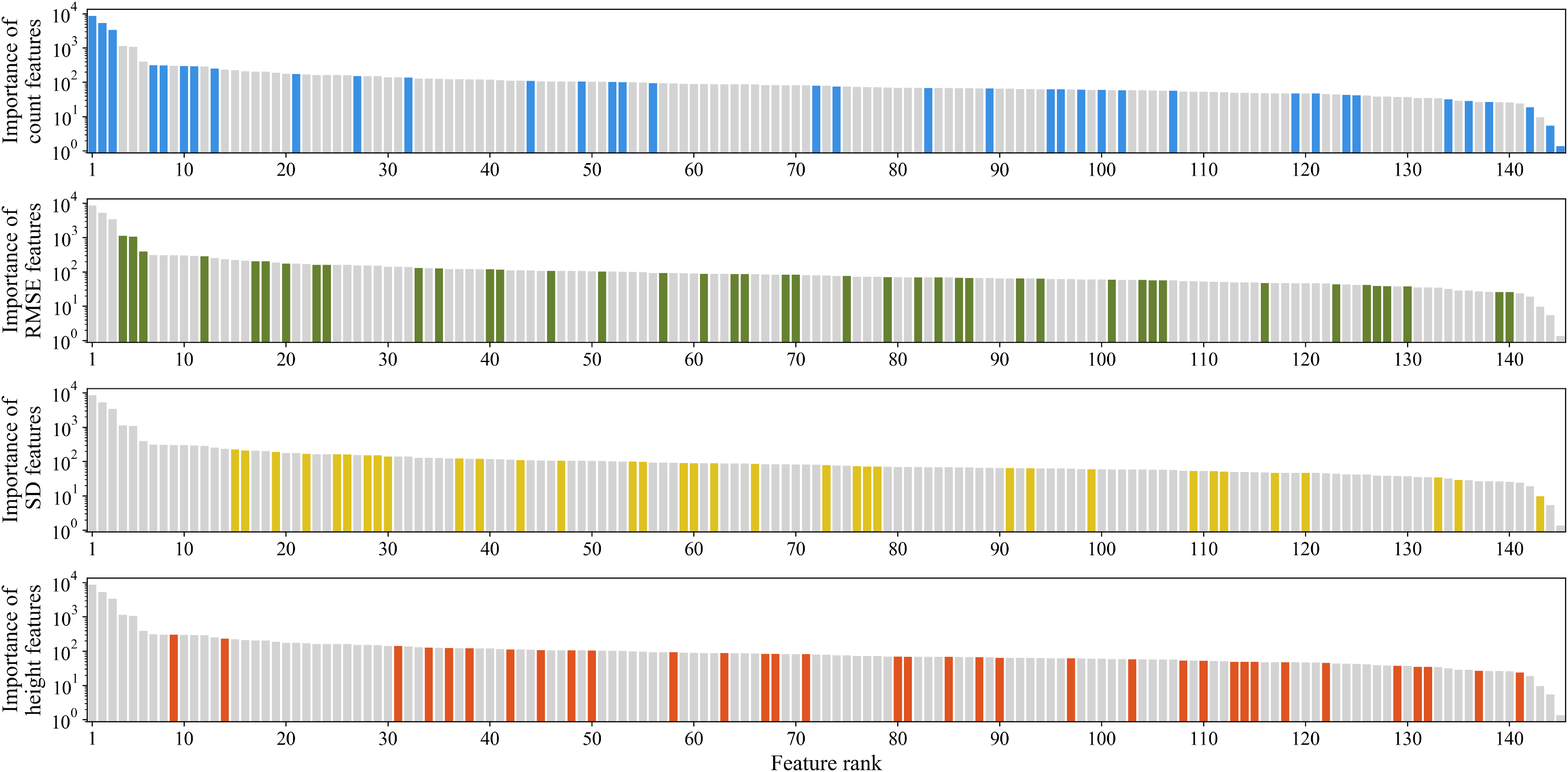}
\caption{The feature importance values of different feature groups provided by the LightGBM model.}
\label{feature_importance}
\end{figure*}

The LightGBM model can evaluate the importance of features by calculating the total gain of each feature. Since the final classification result is given by the combination of multiple LightGBM models, the feature importance of each individual model are added to obtain the final feature importance. Fig. \ref{feature_importance} shows the logarithmic values of feature importance, which are given by the model with an MCC of 0.735. In order to facilitate the comparison of different types of features, they are divided into four categories, and their positions in the feature importance ranking are marked with different colors. The four types of features are count features, RMSE features, SD features and height features. Spefically, RMSE features include both template-matching features and intra-cluster concentration features.

\begin{figure}[!tb]
\centering
\includegraphics[width=8.5cm]{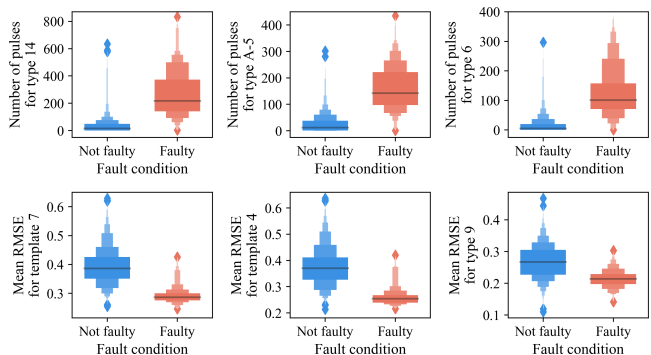}
\caption{Distributions of six features with the highest feature importance values. The top three features are pulse count features and the other three are RMSE features.}
\label{feature_distribution1}
\end{figure}

It can be seen in Fig. \ref{feature_distribution1} that the three most important features are pulse count features, followed by three RMSE features (two of which are template-matching features). The importance values of SD features and average height features are lower overall. The average height features generally have the lowest importance values. It is expected that the distributions of faulty and non-faulty samples are quite different for the most important features. Fig. \ref{feature_distribution1} shows the distributions of faulty and non-faulty samples for the six most important features. The three pulse count features are all cluster-specific, of which the first and third features use pulses from all phases, and the second feature is a single-phase feature (A-5 indicates the fifth cluster of phase A). The fourth and fifth features are template-matching features, and the sixth feature is an intra-cluster concentration feature that uses pulses from three phases. It is observed that the distributions of the top pulse count features of faulty samples are significantly higher than those of non-faulty samples, while the distributions of top RMSE features are significantly lower than those of non-faulty samples. Thus, these features can be used to distinguish the faulty and non-faulty samples.

\begin{figure}[!tb]
\centering
\includegraphics[width=8.5cm]{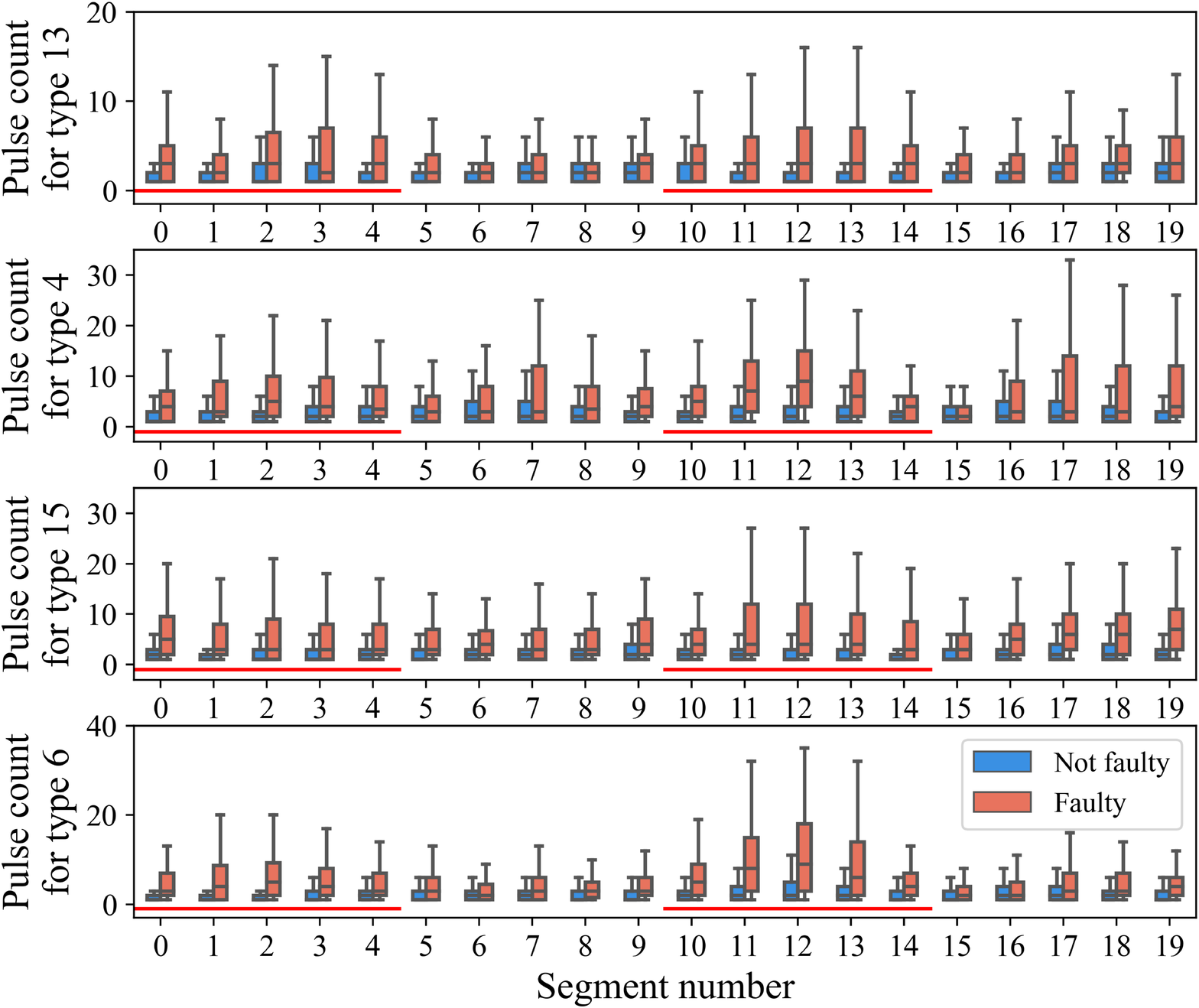}
\caption{Boxplots for phase-resolved distributions of the number of pulses for types 13, 4, 15, and 6 in Fig. \ref{centroids_all} (top to bottom).}
\label{box_Q02}
\end{figure}

A difference between template-matching features and intra-cluster concentration features is that the template-matching features only use pulses within quadrant one and quadrant three of the signals, while the intra-cluster concentration features use all pulses. An obvious reason is that the numbers of pulses used for intra-cluster concentration features are much smaller than template-matching features, which may increase the chance of overfitting if the number of pulses is further reduced. A detailed analysis on the reason that template-matching features use pulses within quadrant one and quadrant three is given in Fig. \ref{box_Q02}, which shows the distributions of pulse counts within phase-resolved segments for the clusters corresponding to the four most important template-matching features. Each signal is evenly divided into 20 signal segments where the pulses are counted separately. Specifically, the four clusters are clusters 13, 4, 15, and 6, which correspond to templates 7, 4, 3, and 1, respectively. For the first two features, the pulse counts in quadrant one and quadrant three (marked with red line) are generally more than quadrant two and quadrant four for faulty samples, while non-faulty samples generally have more pulses in quadrant two and quadrant four. This distributional characteristic indicates that these two features can be more discriminative if the counts are taken for quadrant one and quadrant three only.

\begin{figure}[!tb]
\centering
\includegraphics[width=8.5cm]{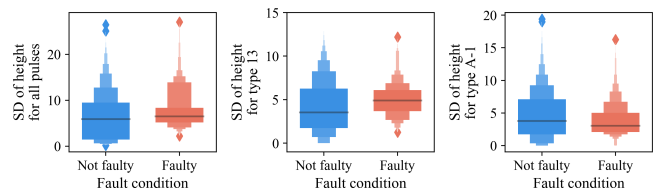}
\caption{Distributions of three SD features with the highest feature importance values.}
\label{feature_distribution2}
\end{figure}

\begin{figure}[!tb]
\centering
\includegraphics[width=8.5cm]{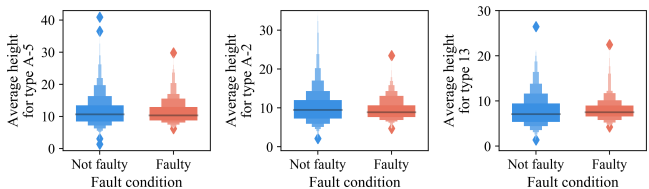}
\caption{Distributions of three average height features with the highest feature importance values.}
\label{feature_distribution3}
\end{figure}

Fig. \ref{feature_distribution2} shows the distributions of three SD features with the highest feature importance, ranking 15, 16, and 19, respectively. Compared with the top-ranked features, the discriminative power of these features for faulty and non-faulty samples has decreased. Nevertheless, the distributions of faulty samples and non-faulty samples for these features are still visibly different. This indicates that these features can still provide useful information during the learning process of DTs.

In Fig. \ref{feature_distribution3}, the distributions of three average height features with the highest feature importance are plotted (the features ranked 9, 14, and 31, respectively). Although the importance ranks of the first two features are relatively high, it can be seen that the distributions of faulty and non-faulty samples are quite similar (almost identical mean values). Compared with other types of features, these features are more likely to lead to overfitting of the models. To justify this argument, we can remove the average height features and evaluate the performance of the model.

\begin{table}[!t]
\renewcommand\arraystretch{1}
\centering  
\captionsetup{justification=centering}
\caption{Comparison of Results for the Proposed Model} 
\begin{tabular}{p{3.5cm} p{1.2cm} p{1.2cm} p{1.2cm}}
\toprule[1.5pt]

Detail & MCC & Precision & Recall\\
 
\midrule[0.75pt]
I 										& 0.735		&  0.669     	&  0.841 	\\
I w/o average height features     		& \textbf{0.740} 	&  \textbf{0.671} 		&  \textbf{0.844}	\\ 
\midrule[0.75pt]
II             							& \textbf{0.750} 	&  0.693  		& 0.838		\\
II (all-pulse features)             	& 0.722 	&  0.615  		& \textbf{0.881}		\\
II (cluster-specific features)          & 0.740 	&  \textbf{0.717}  		& 0.788		\\
\midrule[0.75pt]
III ($\alpha=0.1$)            			& 0.755 	&  0.695  		& 0.846		\\
III ($\alpha=0.15$)             		& \textbf{0.766} 	&  \textbf{0.706}  		& \textbf{0.854}		\\
III ($\alpha=0.2$)          			& 0.751 	&  0.705  		& 0.825	\\

\bottomrule[1.5pt]
\end{tabular}
\label{proposed_comparison}
\end{table}

As mentioned above, the average height features have lower feature importance values in general. Even the features with the highest importance values have lower distinguishability between faulty and non-faulty samples than other feature groups. Therefore, in a new experiment setting, average height features are removed and the LightGBM models are retrained. Next, we change the cross-validation scheme to the standard leave-one-out version and estimate the threshold using validation data, which improves the model's performance (also without average height features). Finally, over-sampling of faulty samples with SMOTE-SVM is implemented in addition to the improved cross-validation scheme. 

A comparison of the results for the proposed approach is presented in Table \ref{proposed_comparison}, where model I represents the model with all features, model II represents the model with the improved cross-validation scheme, and model III represents model II with SMOTE-SVM applied to each split of cross-validation. For model II, additional results are added for models with only all-pulse features or cluster-specific features. As shown in the table, the average height features have a negative impact on the performance of the classification model when other groups of features are used. Shifting to the standard leave-one-out cross-validation also improves the model's performance. Over-sampling the faulty samples with SMOTE-SVM with an $\alpha$ between 0.1 and 0.2 generally lifts the MCC, and model III with $\alpha=0.15$ has the highest MCC. Nevertheless, as a single threshold is used for the evaluation of the metrics, the trade-off between precision and recall leads to the discussion in the following subsection.

\subsection{Performance Comparison With Varying Thresholds}

\begin{figure*}[!tb]
\centering
\includegraphics[width=14cm]{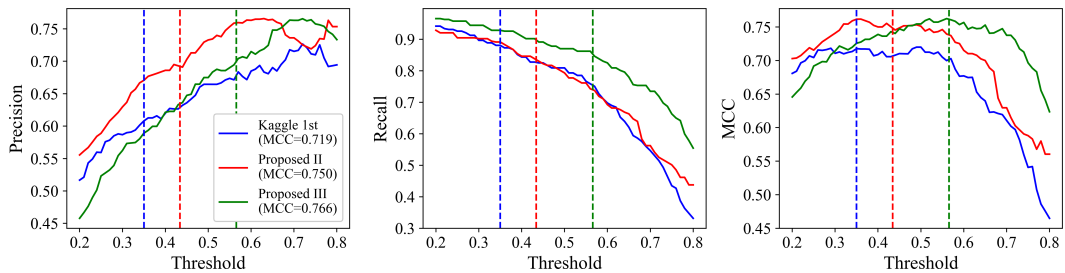}
\caption{Comparison between the proposed model and the 1st place solution on Kaggle with varying thresholds.}
\label{threshold}
\end{figure*}

For each test sample, the classification model first gives the sample a ``fault probability’’ between 0 and 1, and then determines whether the sample is faulty or not based on a threshold obtained by cross-validation. In fact, as the threshold moves from 0 to 1, the values of both TP and FP gradually decrease, while the value of FN gradually increases. Therefore, the recall value decreases monotonically with the increase of the threshold. The precision value, whose shape is determined by the distribution of the output probabilities of the samples, tends to increase with the increase of the threshold. In order to have a clear understanding of the performance of different models, we draw curves of various metrics obtained by varying the threshold within a wide range. 

Fig. \ref{threshold} shows a comparison of the proposed model and the 1st place solution on Kaggle for several metrics when the threshold varies from 0.2 to 0.8. As the figure suggests, the proposed model II and Kaggle's 1st place solution have similar recall values, but model II has a significant improvement on the precision value. In fact, the threshold estimated for model II is 0.434, while the threshold for the 1st place solution on Kaggle is 0.350. The result shows, however, that model II performs better on MCC for all thresholds ranging from 0.2 to 0.8. The proposed model III significantly improves the recall metric with an estimated threshold of 0.566. As expected, over-sampling the faulty samples is prone to producing more false positive samples and fewer false negative samples. In summary, model III has the highest MCC at its estimated threshold and outperforms other models when the threshold is greater than 0.5.

\subsection{Discussion on the Number of Clusters}

In Section \ref{section:sec3.2}, it is mentioned that the number of clusters can be determined by two methods, namely, the heuristic method and the method based on SSE evaluation. Although pulse waveform clustering is only an intermediate process of feature construction, the analysis in Section \ref{section:sec3.2} implies that a proper choice of the number of clusters can help reveal the pulse patterns. 

Specifically, increasing the number of clusters results in a more finely-divided space of pulse waveforms, so that the constructed features can capture the distribution of pulse waveform data more precisely. However, a large number of clusters reduces the number of pulse waveforms within each cluster, thus it is more likely to cause the problem of overfitting. Therefore, it is necessary to examine the effect of the number of clusters on the model's performance. We conduct experiments with different number of clusters and compare the MCC values of each case. Model II with only cluster-specific features is used to have a rigorous comparison.

\begin{figure}[!tb]
\centering
\includegraphics[width=6cm]{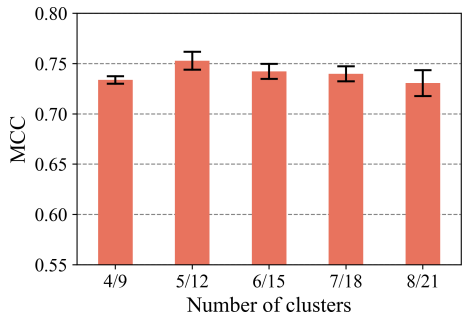}
\caption{The MCC values of the proposed model using cluster-specific features with different cluster numbers. Each value is averaged from five trials and the error bars indicate the range of one SD above and below the average values.}
\label{num_clusters}
\end{figure}

The MCC values of the proposed model with different cluster numbers are shown in Fig. \ref{num_clusters}. The labels for different cases represent the numbers of clusters for pulses from individual phases and all phases, respectively. Five trials are conducted for each case (clustering results differs from trial to trial as the pulses used for clustering are randomly sampled) and the SD values are calculated to yield the error bars. With only cluster-specific features available, the proposed model is able to achieve MCC values larger than 0.73 for all cases. Two conclusions can be drawn from Fig. \ref{num_clusters}:

\begin{itemize}
\item The number of clusters has a certain influence on the performance of the model. However, Fig. \ref{num_clusters} shows that the performance does not vary much for different numbers of cluster within a certain range. Nevertheless, the methods of determining the number of clusters provided in Section \ref{section:sec3.2} are able to find the numbers that result in relatively high MCC values. 
\item Increasing the number of clusters will increase the variance of the MCC values. This is probably due to the fact that the number of pulses assigned to each cluster decreases, so the model is more prone to overfitting.
\end{itemize}

\section{Conclusion}

In this paper, we propose an efficient approach for faulty and non-faulty covered conductor classification based on PD patterns. With a meticulous design of data pre-processing procedure, pulses related to PD patterns are revealed. In-depth analysis of pulse patterns and clustering of pulse waveforms enable the construction of cluster-specific and all-pulse features. In addition to traditional features, two new features, namely, template-matching degree, and intra-cluster concentration degree are proposed. An in-depth analysis of feature importance reveals that average height features can be excluded for the fault detection task. Over-sampling the faulty data samples near the borderline is proved to be useful in dealing with the imbalance of faulty and non-faulty samples. Extensive results show that the proposed approach, which achieves an MCC value of 0.766, has a significant edge over existing methods. Results with varying classification thresholds further corroborate the superiority of the proposed approach. Furthermore, our generic design can be applied to many other pattern recognition tasks.

Our future goal will be focused on performance enhancement by using deep neural networks. Deeper evaluation of representative pulse waveforms can be utilized to identify whether they come from noise interference, PD effect or other sources. Such a procedure may further improve the explainability of the proposed approach. 

{
\small
\bibliographystyle{IEEEtran}%
\bibliography{IEEE.bib}
}

\begin{IEEEbiography} [{\includegraphics[width=1in,height=1.25in,clip,keepaspectratio]{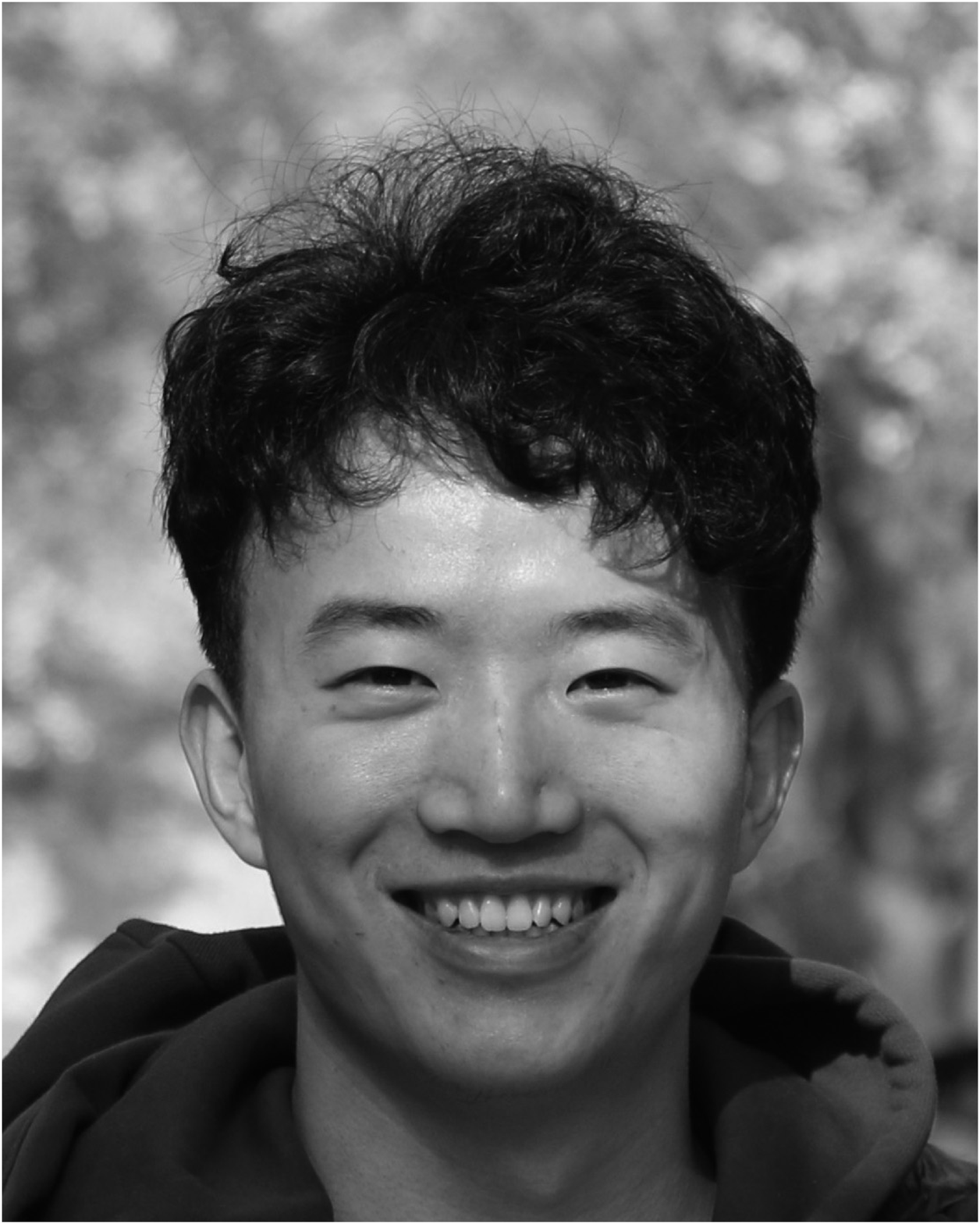}}]
{Kunjin Chen} received his B.Sc. and Ph.D. degrees in electrical engineering from the Department of Electrical Engineering, Tsinghua University, Beijing, China, in 2015 and 2020.

His research interests include a variety of machine learning applications in power systems.
\end{IEEEbiography}


\begin{IEEEbiography} [{\includegraphics[width=1in,height=1.25in,clip, keepaspectratio]{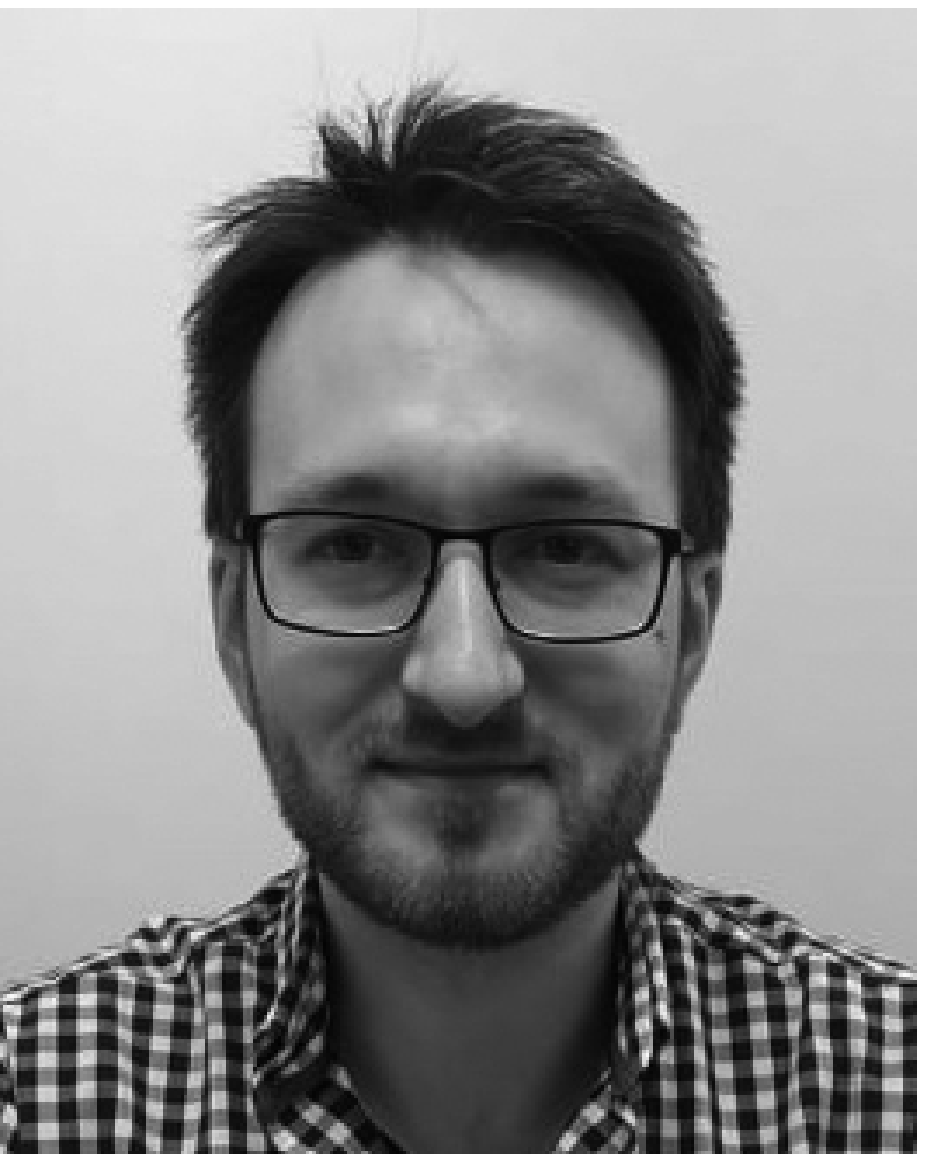}}]
{Tom\'{a}\v{s} Vantuch} received his PhD degree in Computer Science in 2018 at V\v{S}B - Technical University of Ostrava, Czech Republic. He has published a number of articles in peer reviewed journals and conference proceedings. He actively participates in several national and international projects.

His current work includes implementation of smart grid technologies using prediction models and bio-inspired methods as well as utilization of machine learning models for industrial applications.
\end{IEEEbiography}


\begin{IEEEbiography} [{\includegraphics[width=1in,height=1.25in,clip,keepaspectratio]{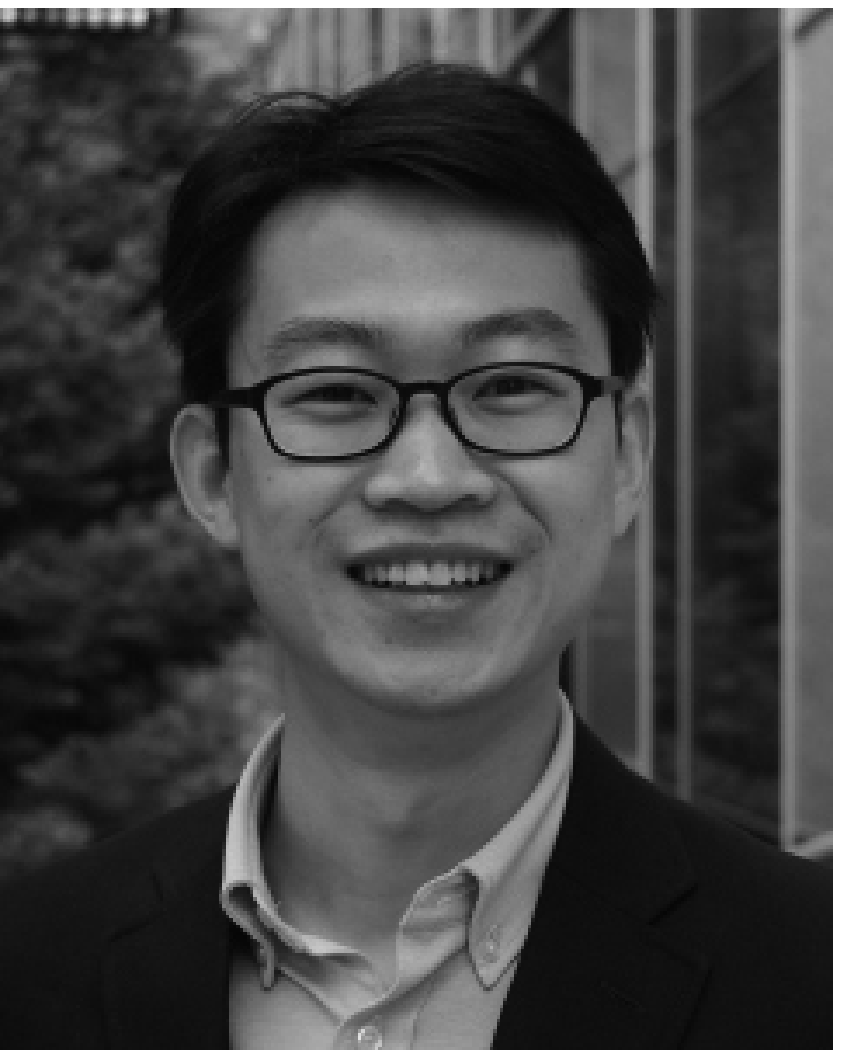}}]
{Yu Zhang} (M'15) received the Ph.D. degree in electrical and computer engineering from the University of Minnesota, Minneapolis, MN, USA, in 2015.

He is an Assistant Professor in the ECE Department of UC Santa Cruz. Prior to joining UCSC, he was a postdoc at UC Berkeley and Lawrence Berkeley National Laboratory. His research interests span the broad areas of cyber-physical systems, smart power grids, optimization theory, machine learning and big data analytics.
\end{IEEEbiography}


\begin{IEEEbiography} [{\includegraphics[width=1in,height=1.25in,clip,keepaspectratio]{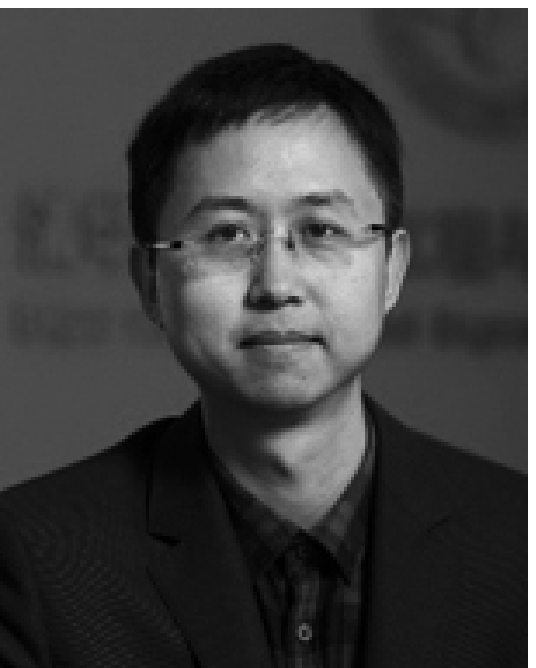}}]
{Jun Hu} (M'10) received his B.Sc., M.Sc., and Ph.D. degrees in electrical engineering from the Department of Electrical Engineering, Tsinghua University in Beijing, China, in July 1998, July 2000, July 2008. 

Currently, he is an associate professor in the same department. His research fields include overvoltage analysis in power system, sensors and big data, dielectric materials and surge arrester technology.
\end{IEEEbiography}

\vspace{-325pt}

\begin{IEEEbiography} [{\includegraphics[width=1in,height=1.25in,clip,keepaspectratio]{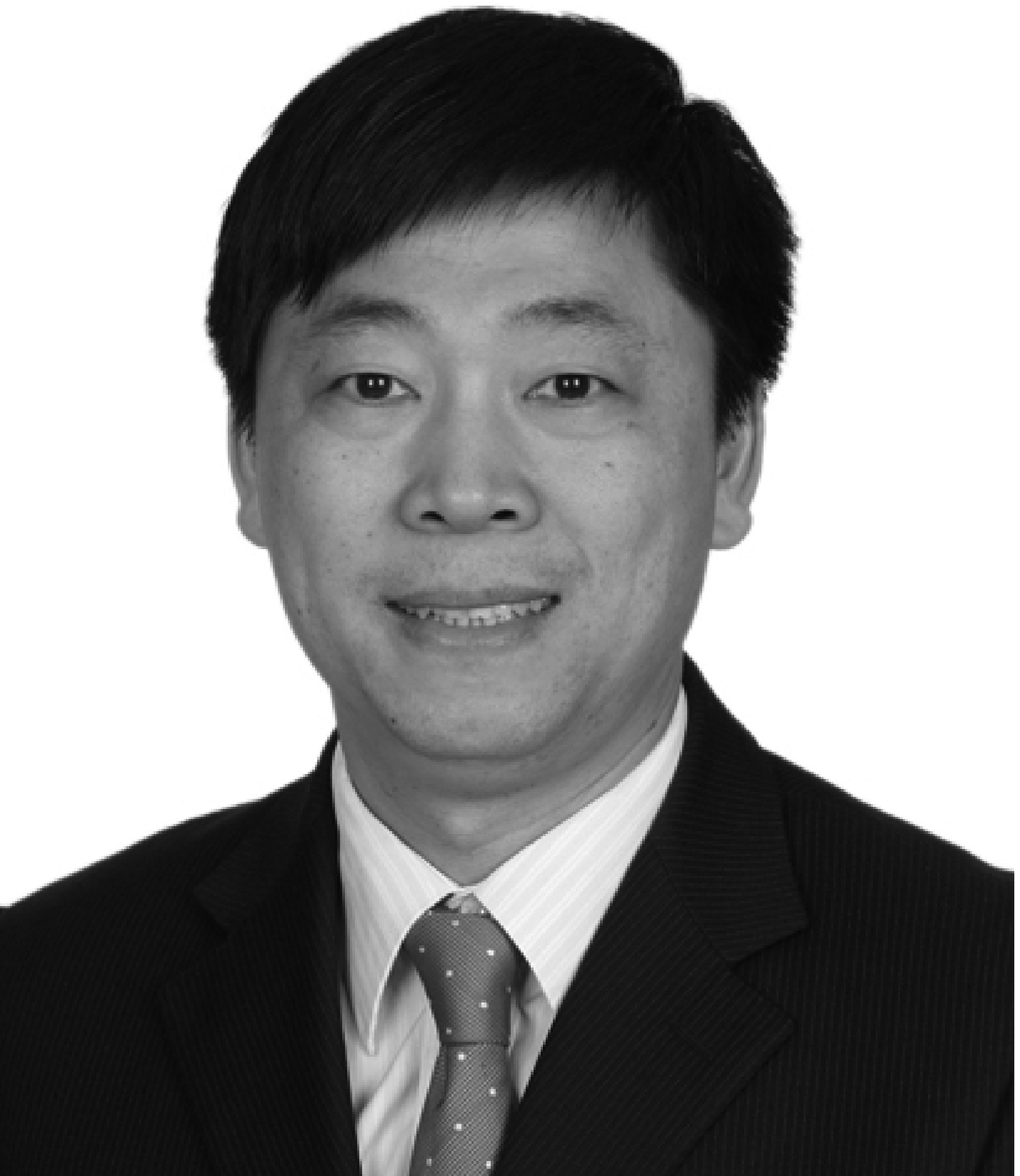}}]
{Jinliang He} (M'02--SM'02--F'08) received the B.Sc. degree from Wuhan University of Hydraulic and Electrical Engineering, Wuhan, China, the M.Sc. degree from Chongqing University, Chongqing, China, and the Ph.D. degree from Tsinghua University, Beijing, China, all in electrical engineering, in 1988, 1991 and 1994, respectively.

He became a Lecturer in 1994, and an Associate Professor in 1996, with the Department of Electrical Engineering, Tsinghua University. From 1997 to 1998, he was a Visiting Scientist with Korea Electrotechnology Research Institute, Changwon, South Korea, involved in research on metal oxide varistors and high voltage polymeric metal oxide surge arresters. From 2014 to 2015, he was a Visiting Professor with the Department of Electrical Engineering, Stanford University, Palo Alto, CA, USA. In 2001, he was promoted to a Professor with Tsinghua University. He is currently the Chair with High Voltage Research Institute, Tsinghua University. He has authored five books and 400 technical papers. His research interests include overvoltages and EMC in power systems and electronic systems, lightning protection, grounding technology, power apparatus, and dielectric material.
\end{IEEEbiography}

\end{document}